\documentclass[fleqn,usenatbib]{mnras}

\usepackage{newtxtext,newtxmath}

\usepackage[T1]{fontenc}

\DeclareRobustCommand{\VAN}[3]{#2}
\let\VANthebibliography\thebibliography
\def\thebibliography{\DeclareRobustCommand{\VAN}[3]{##3}\VANthebibliography}

\usepackage{graphicx}	
\usepackage{amsmath}	
\usepackage{tabularx}

\usepackage{xspace}
\usepackage{comment}

\newcommand{\NHI}{\relax\ifmmode{N_{\rm HI}\xspace} \else {$N_{\rm HI}$}\expandafter\xspace\fi}
\newcommand{\unitNHI}{\ifmmode{\rm \,cm^{-2}}\else\,cm$^{-2}$\xspace\fi}  
\newcommand\Ha{H$\alpha$}
\newcommand\Hb{H$\beta$}
\newcommand\Hg{H$\gamma$}
\newcommand\HI{H~I}

\newcommand\OIII{[O~III]}
\newcommand\NII{[N~II]}
\newcommand\OVI{O~VI}
\newcommand\HeII{He~II}

\title[Raman He~II Lines in RR Tel]{High-Resolution Spectroscopy of Raman-scattered He~II Lines in the Symbiotic Nova RR~Telescopii}

\author[Shin, Chang et al.]{
Jaejin Shin,$^{1}$
Seok-Jun Chang,$^{2}$
\thanks{Corresponding author}\thanks{Email: sjchang@mpa-garching.mpg.de}
Hee-Won Lee,$^{1}$
Sam Kim,$^{3}$
\\
$^{1}$ Department of Physics and Astronomy, Sejong University, Seoul, Republic of Korea \\
$^{2}$ Max-Planck-Institut f\"{u}r Astrophysik, Karl-Schwarzschild-Stra$\beta$e 1, 85748 Garching b. M\"{u}nchen, Germany \\
$^{3}$ ESO, Alonso de Córdova 3107, Vitacura, Chile
}

\date{Accepted XXX. Received YYY; in original form ZZZ}

\pubyear{\the\year{}}

\begin{document}
\label{firstpage}
\pagerange{\pageref{firstpage}--\pageref{lastpage}}
\maketitle

\begin{abstract}
Raman-scattered emission features in symbiotic stars provide a powerful diagnostic of mass-loss and transfer processes, as they uniquely probe both ionized and neutral regions within interacting binaries. When resolved with high-resolution spectroscopy, these features encode detailed information on the physical properties of the neutral hydrogen medium.
In this work, we present high-resolution spectroscopic observations of the symbiotic nova RR Telescopii obtained with FEROS in 2004 and GHOST in 2024, providing a $\sim$20 yr baseline. We report the clear detection of all three Raman-scattered He~II lines at 6545 \AA, 4851 \AA, and 4332 \AA, and constrain the distribution and kinematics of H~I through line profile analysis. The three Raman lines exhibit distinct relative velocities, indicating that they trace different depths within the H~I region. The Raman conversion efficiencies of the three Raman He~II lines in 2024 are significantly lower than those in 2004, 
indicating substantial changes in the physical properties of the neutral hydrogen region. In addition, radiative transfer modeling implies a larger covering factor (opening angle) of the neutral region in 2004 than in 2024.
These results indicate that the neutral hydrogen region cannot be characterized by a single H~I column density, emphasizing the need for advanced radiative transfer modeling that accounts for the complex kinematics and geometry of the H~I region.
Overall, these results establish Raman-scattered He II lines as a powerful tool for spectroscopic tomography, allowing for direct constraints on the structure and kinematics of neutral hydrogen in symbiotic binaries.
\end{abstract}




\section{Introduction}\label{sec:intro}
Symbiotic stars are interacting binary stellar systems consisting of a white dwarf (WD) and a red giant (RG). These systems display a range of astrophysical phenomena, including outbursts, prominent ultraviolet (UV) and optical emission lines, collimated outflows, and X-ray emission \citep{mikolajewska12,Munari2019,Merc2025}. These features result from accretion through the gravitational capture of the slow stellar wind emanating from the giant companion. Consequently, symbiotic stars serve as natural laboratories for studying mass transfer, accretion physics, and binary evolution during the late stages of stellar life \citep[e.g.,][]{kenyon86, devalborro17, skopal23}.

Based on their infrared (IR) spectral energy distributions, symbiotic stars are classified into two types: S-type and D-type. D-type symbiotic stars exhibit an IR excess, indicative of warm dust, whereas S-type systems do not show such an excess \citep{Ivison95,angeloni10, jurkic12,akras19}. The orbital parameters of D-type binaries are generally poorly constrained \citep{belczynski00,akras19}, in contrast to those of S-type systems, which typically have orbital periods of several hundred days \citep[e.g.,][]{mikolajewska12}. Notably, the mass donor in D-type symbiotic stars is usually a Mira-type variable or an OH/IR source \citep[e.g.,][]{cho10}. 

A key observational signature of about half of symbiotic stars is the presence of broad emission features resulting from Raman-scattering \citep{akras19,merc19a,merc19b}, which directly probes the neutral hydrogen surrounding the binary system. Early work by \cite{schmid89} identified two prominent broad features at 6830 and 7088 \AA, produced by Raman-scattering of the \OVI\ resonance doublet at $\lambda\lambda1032,1038$ by atomic hydrogen. Subsequent studies revealed additional, weaker Raman-scattered lines of \HeII\, with far-UV transitions at $\lambda\lambda$1025, 972, and 949 being scattered into optical features at 6545, 4851, and 4332 \AA\ \citep{vangroningen93,lee12}.

These Raman-scattered emission lines provide a direct probe of atomic hydrogen during the mass loss and transfer processes in symbiotic stars. In particular, Raman-scattered \HeII\ lines are especially useful because higher-level Raman-scattered \HeII\ lines have smaller cross sections, enabling the probing of neutral regions with higher column densities. 
Intriguingly, Raman-scattered \HeII\ emission lines have thus far been detected only in D-type symbiotic stars (i.e., RR Tel, V1016 Cyg, and HM Sge; \citealt{vangroningen93,lee01,lee03,birriel04,lee12}). Among these, RR Telescopii (RR Tel) is of particular interest. RR~Tel is a well-known symbiotic nova that underwent a nova-like outburst in 1944 and has been fading ever since \citep[e.g.,][]{mayall49}. According to the American Association of Variable Star Observers (AAVSO), RR~Tel was observed to be $\sim 6$ mag in 1948 and steadily dimmed to $\sim 12$ mag by 2024. Characterized by a Mira-type giant as its mass donor, RR~Tel exhibits a pronounced infrared excess consistent with its classification as a D-type symbiotic star. It also displays a remarkably rich spectrum of emission lines spanning a wide range of ionization and excitation states \citep{selvelli00}.

In addition, RR~Tel is an excellent object for investigating Raman-scattered \HeII. Notably, RR~Tel exhibits Raman-scattered \HeII\ at 6545, 4851, and even 4332 \AA\ \citep{vangroningen93, lee01}. Because these scattered spectral features probe different parts of the neutral region near the giant star, they provide crucial information about the mass-loss process during the late stages of stellar evolution. However, no quantitative investigation of these scattered features has been carried out. Therefore, a detailed study of Raman-scattered \HeII\ is necessary to better understand the nature and physics of the neutral region in symbiotic stars.

For this purpose, we investigate Raman-scattered \HeII\ in RR~Tel using high-resolution spectroscopic data obtained with Gemini/GHOST and MPG/FEROS. In Section~\ref{sec:atomic}, basic atomic physics is summarized.
Section~\ref{sec:spectroscopy} describes the observations and data preparation, followed by data analysis in Section 4. The main results, with their theoretical interpretation, are presented in Section~5. Finally, the discussion and summary are provided in Sections~6 and 7.\\

\begin{figure}
    \centering
    \includegraphics[width=0.97\linewidth]{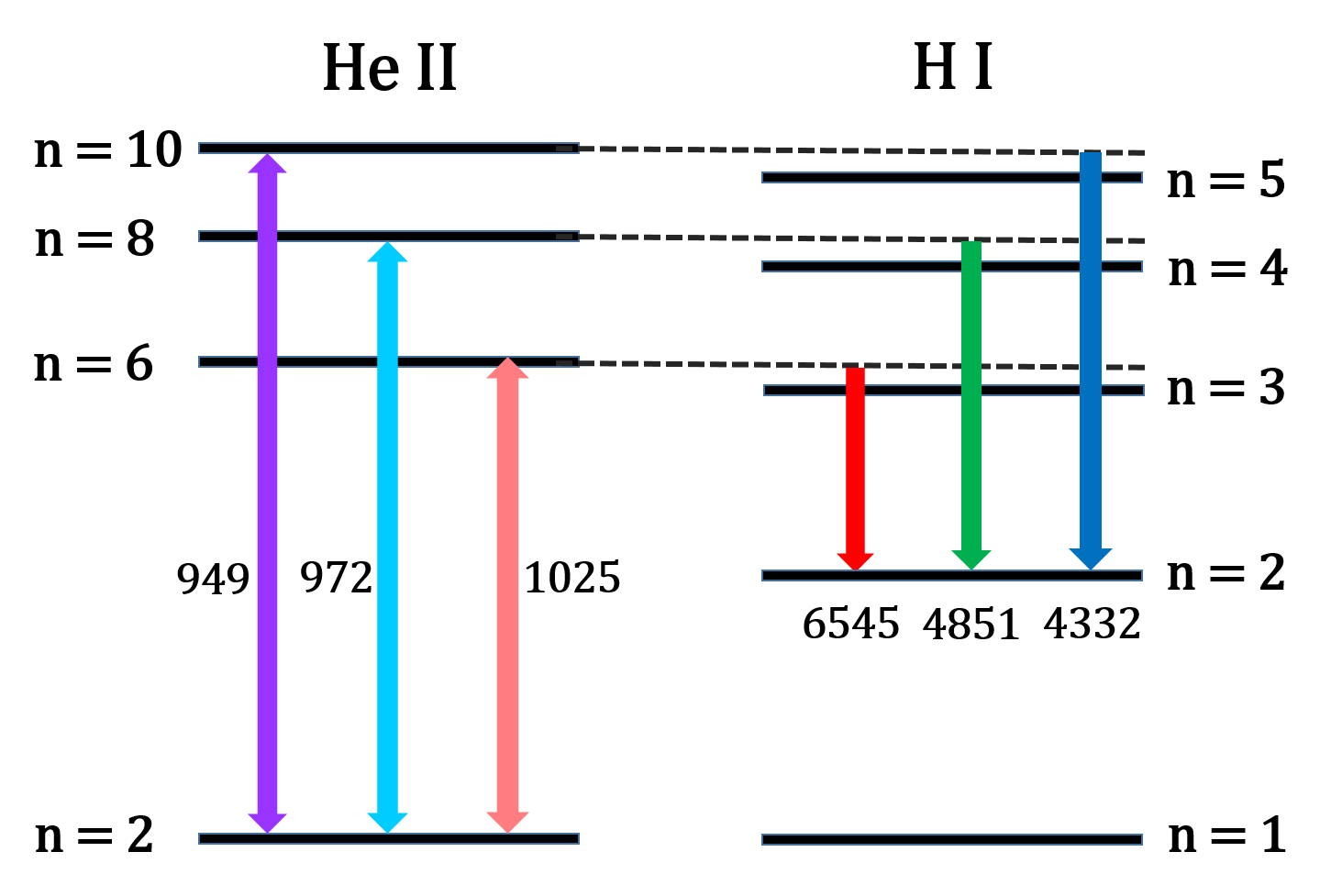}

    \caption{Schematic illustration of \HeII\ atomic transitions and the formation of Raman-scattered \HeII\ lines near H$\alpha$, H$\beta$, and H$\gamma$. Far-UV \HeII\ Balmer lines near the hydrogen Lyman series can excite the ground-state electron, and subsequent de-excitation to the n = 2 state emits Raman-scattered He~II lines blueward of the hydrogen Balmer lines.}
    \label{fig:atomic}
\end{figure}

 \begin{table}
 \centering
 \caption{Scattering cross sections and wavelengths of three far UV \HeII Balmer lines for Rayleigh scattering into $n=1$ ($\sigma_{n = 1, \rm Ray}$) and Raman-scattering into $n=2$ ($\sigma_{n = 2, \rm Ram}$) }
  \begin{tabular}{ccccc}
 \hline
    Transition    &  $6\to2$   & $8\to 2$  &  $10\to 2$   \cr    
 \hline
 He~II$\lambda$ (\AA) & 1025.28    &   972.13     &  949.32      \cr
 $\sigma_{n = 2, \rm Ram}\ (\rm cm^2)$   & $1.2\times10^{-21}$ & $1.4\times10^{-22}$    &  $2.9\times10^{-23}$     \cr
 $\sigma_{n = 1, \rm Ray}\ (\rm cm^2)$  &  $6.2\times10^{-21}$ & $8.3\times10^{-22}$     & $1.9\times10^{-22}$      \cr
Raman He~II$\lambda$ (\AA) &  6544.70   &  4851.30    & 4331.74     \cr

 \hline
  \end{tabular}
  \label{tab:atomic}
 \end{table}

\section{Basic Atomic Physics}\label{sec:atomic}

Far-UV He~II Balmer lines associated with the $2n\rightarrow 2$ transitions have slightly higher energies than the hydrogen Lyman series ($n\rightarrow 1$). 
These He~II Balmer lines can excite an electron in the ground state of the atomic hydrogen to $n > 2$ states. When the excited electron de-excites to an intermediate state ($n \geq 2$), the inelastic scattering process is referred to as Raman scattering. In contrast, elastic scattering to the ground state is known as Rayleigh scattering.
Therefore, He~II Balmer lines are efficiently Raman-scattered with atomic hydrogen, generating optical emission features that appear near the hydrogen Balmer lines \citep[e.g.,][]{nussbaumer89, vangroningen93, lee01}. Figure~\ref{fig:atomic} shows a schematic diagram of the first three transitions ($n=3,\ 4,\ 5$), which produce Raman-scattered He~II features located blueward of H$\alpha$, H$\beta$, and H$\gamma$.

The interaction between photons and bound electrons is treated using second-order time-dependent perturbation theory, as outlined in standard quantum mechanics references \citep[e.g.,][]{bethe67, sakurai67}. In particular, the cross sections for Rayleigh and Raman scattering of far-UV photons by neutral hydrogen are described by the Kramers–Heisenberg formula.
Table~\ref{tab:atomic} summarizes the cross sections and central wavelengths of the He~II Balmer lines, along with the corresponding Raman-scattered features blueward of the hydrogen Balmer lines for $n=3,\ 4,\ 5$ \citep[e.g.,][]{lee12, chang23, kokubo24}.
Note that the wavelengths of far-UV He~II and optical Raman He~II are expressed in vacuum and air, respectively.

By energy conservation, the wavelength $\lambda_{\rm Raman}$ of a Raman-scattered line is connected to the incident photon wavelength, $\lambda_{\rm inc}$, by
\begin{equation}
\lambda_{\rm inc}^{-1} = \lambda_{\rm Ram}^{-1} + \lambda_{\rm Ly\alpha}^{-1},
\label{eq:wavelength}
\end{equation}
where $\lambda_{\rm Ly\alpha}$ denotes the hydrogen Ly$\alpha$ wavelength. From Equation~(\ref{eq:wavelength}), we obtain
\begin{equation}
\frac{\Delta \lambda_{\rm inc}}{\lambda_{\rm inc}} = \left( \frac{\lambda_{\rm inc}}{\lambda_{\rm Ram}} \right) \frac{\Delta \lambda_{\rm Ram}}{\lambda_{\rm Ram}}.
\label{eq:linewidth}
\end{equation}
Thus, the inelastic nature of Raman-scattering broadens spectral features by a factor of $\lambda_{\rm Raman}/\lambda_{\rm inc}$. These Raman-scattered He~II features appear as weak, broadened components near hydrogen Balmer lines.



\section{Observation and Data} \label{sec:spectroscopy}

We performed a high-resolution, deep spectroscopic analysis to investigate Raman-scattered \HeII\ in RR~Tel based on data obtained with the Gemini High-resolution Optical Spectrograph (GHOST) and the Fiber-fed Extended Range Optical Spectrograph (FEROS). These datasets provide a nearly 20-year baseline for RR~Tel. 

\subsection{Spectroscopy with GHOST}
The GHOST observations were carried out on March 25, 2024 (Program ID: GS-2024A-Q-305; PI: Hee-Won Lee). GHOST covers the wavelength range from 3,830 to 10,000 \AA\ with a spectral resolution of R$\sim$56,000. The observation was conducted in the ``standard-resolution" mode. Because Raman-scattered He~II features are very faint (e.g., the line flux of Raman-scattered \HeII\ at 6545 \AA\ is approximately 1/2000 that of \Ha, see \citealt{lim25}), 2$\times$2 binning was employed to maximize the signal-to-noise ratio in the continuum. With this binning, the effective spectral resolution achieved was approximately R$\sim$40,000.

To cover both strong (e.g., \Ha) and weak (e.g., Raman-scattered \HeII) spectral lines, we obtained spectra of RR Tel with various exposure times: 10, 100, 500, and 1,000 seconds. The shorter exposures (10 and 100 seconds) targeted strong emission lines, such as [O III] and \Ha, while the longer exposures (500 and 1,000 seconds) were designed to detect the Raman-scattered \HeII\ line. We note that \Ha\ was saturated in these longer exposures. The total on-source exposure time was 3,230 seconds.

We retrieved reduced data from the Gemini archive \citep{Placco2024} and combined the spectra. Since the \Ha\ line was saturated in the longer exposures, we combined only the short exposures for the \Ha\ window.

\subsection{Spectroscopy with FEROS}
To complement our recent GHOST data and establish a long-term baseline, we analyzed archival high-resolution spectra obtained with FEROS. FEROS is a highly efficient echelle spectrograph mounted on the MPG 2.2m telescope located at La Silla Observatory. The instrument provides a spectral resolution of R$\sim$48,000 and covers 3,500--9,200 \AA. 

The specific FEROS data utilized were retrieved from the  ESO Science Archive (Program ID: 073.D-0724, PI: R. Zamanov). The observations were performed on two separate dates: June 4 and August 30,  2004, thereby providing a $\sim$20-year baseline relative to our GHOST data. For the analysis, four individual exposures, each with a 600 s integration time, were co-added to maximize the signal-to-noise ratio. Since FEROS spectra are not flux-calibrated, we used these data to measure flux ratios between the Raman-scattered \HeII\ features and their nearby intrinsic \HeII\ emission lines. This ensures that our analysis is largely unaffected by the lack of absolute flux calibration.

\section{Analysis} \label{sec:floats}
\subsection{Line Fitting Procedure}
In this work, we performed a multi-component line-fitting analysis of the GHOST and FEROS spectra for RR~Tel. Given the large number of emission lines present throughout the spectrum, we conducted separate fitting procedures for \Ha, \Hb, and \Hg\ windows to minimize interference from unrelated emission or absorption features.

First, for each spectral window (\Ha, \Hb, and \Hg), we fitted and subtracted a local continuum determined from the emission-free regions on either side of the lines of interest. We then fitted the emission lines in the residual spectrum, including the Balmer lines, \HeII, prominent forbidden lines (e.g., [O~III], [N~II]), and the Raman-scattered \HeII\ lines.

During the fitting process, we adopted a Gaussian profile for all emission lines except the Balmer lines. For the Balmer lines, we identified prominent wing features, previously reported in symbiotic stars \citep{nussbaumer89, lee00, chang18}. Because of these wing features and their asymmetric line profiles, we applied a double Voigt profile to reproduce the observations. Additionally, several constraints were imposed during the fitting process. First, we tied the kinematics of \NII$\lambda$6548 and \NII$\lambda$6583, fixing their flux ratio at 1:3, which corresponds to the theoretical branching ratio of this forbidden doublet. Second, we constrained the velocity dispersion ratio between Raman-scattered \HeII\ and \HeII\ lines to account for the line broadening of Raman-scattered \HeII\ emission \citep[see Sect. 2 of][]{lim25}. For Raman-scattered \HeII\ at 6545, 4851, and 4332~\AA, the broadening factors are 6.38 (6544.70/1025.28), 4.99 (4851.30/972.13), and 4.56 (4331.74/949.32), respectively.

\begin{figure*}
    \centering
    \includegraphics[width=\linewidth]{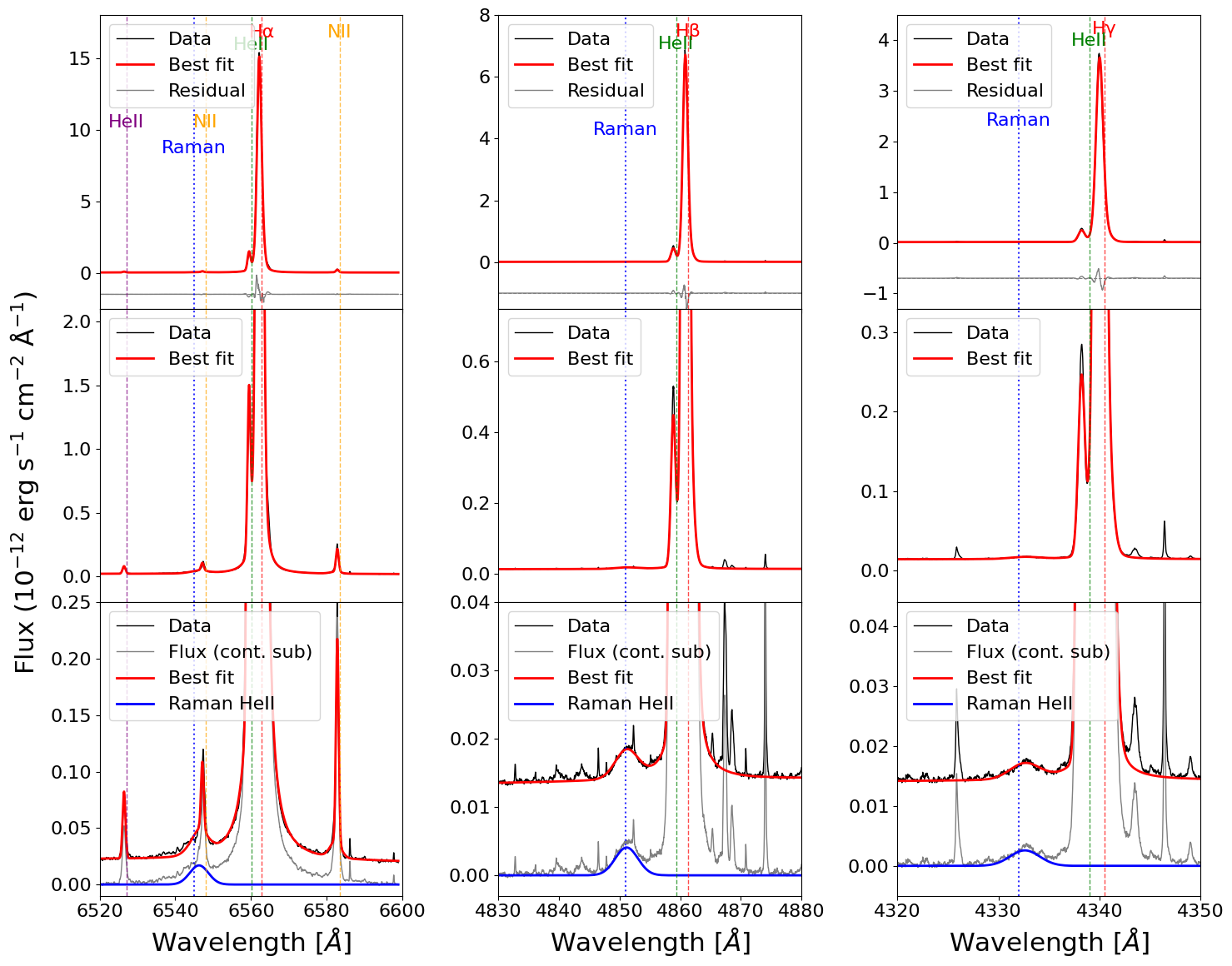}
    \caption{ Fitting results of the GHOST data. From left to right, the \Ha, \Hb, and \Hg\ spectral windows are presented. The top, middle, and bottom panels show the full flux range, a $\sim$10-times magnified view, and a $\sim$100--200-times magnified view, respectively. The black, red, and blue lines represent the data, the best-fit model, and the Raman-scattered \HeII, respectively. The gray lines indicate residuals in the top panel and the continuum-subtracted flux in the bottom panel. The Raman-scattered \HeII\ emission lines are clearly visible in the bottom panels.
    }
    \label{fig:GHOST}
\end{figure*}

\begin{figure*}
    \centering
    \includegraphics[width=\linewidth]{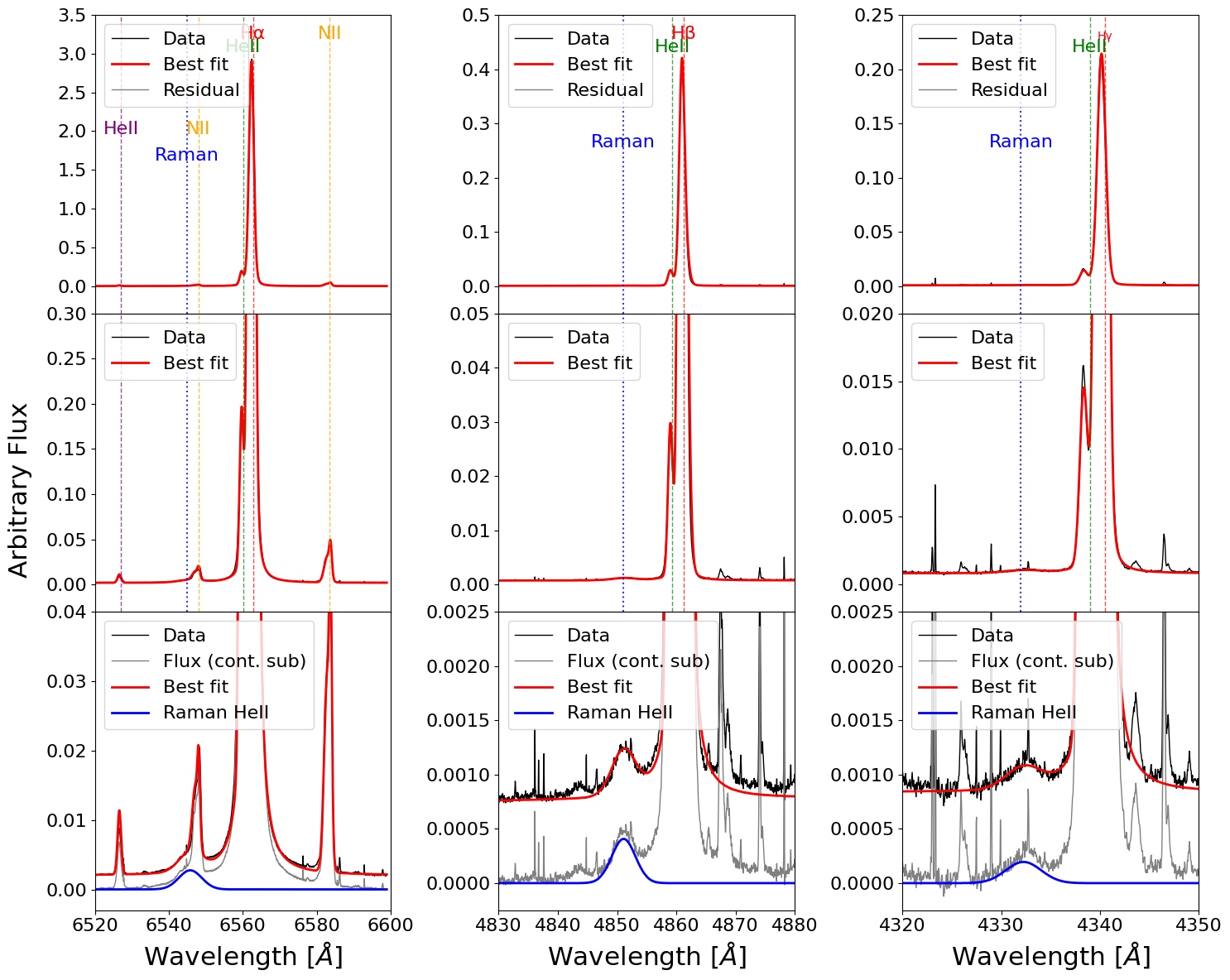}
    \caption{Fitting results of the FEROS data. Other information is the same as in Figure~\ref{fig:GHOST}.}
    \label{fig:FEROS}
\end{figure*}

We show the fitting results of the GHOST and FEROS data in Figure~\ref{fig:GHOST} and Figure~\ref{fig:FEROS}, respectively. In both figures, strong Balmer lines and \HeII\ lines are readily apparent in the upper panels. The middle panels show spectra magnified by about a factor of 10, clearly displaying subordinate optical \HeII\ lines such as He~II$\lambda$6560. The lower panels, magnified by several hundred times, reveal weaker emission lines, including Raman-scattered features. As shown in the lower panels, Raman-scattered \HeII\ lines at 6545, 4851, and 4332 \AA\ are clearly detected and fitted with Gaussian profiles.

The best-fit model parameters for the emission lines, including the Raman-scattered features, are listed in Table~\ref{tab:FIT}. 

\begin{table}
\centering
\caption{Fitting Results of Raman scattered He~II and optical He~II lines using GHOST and FEROS data. 
The columns, from left to right, represent line types, line fluxes, central wavelengths, and line widths of \HeII\ and Raman-scattered \HeII.
}
\begin{tabular}{cccc}
\hline
\multicolumn{4}{c}{\textbf{GHOST}}	  \cr
\hline
Lines	&	Flux 	&	Center 	&	$\sigma$ 	\cr
& ($10^{-14}$ erg s$^{-1}$ cm$^{-2}$)
 & (\AA) & (\AA) \cr
\hline
HeII6560	&	120.06 	&	6559.46 	&	0.39 	\cr
Raman 6545	& 10.49 	&	6546.27 	&	2.47 	\cr
HeII4859	&	39.07 	&	4858.84 	&	0.38 	\cr
Raman 4851	&	1.91 	&	4851.14 	&	1.87 	\cr
HeII 4338	&	18.03	&	4338.23 	&	0.32 	\cr
Raman 4332	&	0.95 	&	4332.61 	&	1.47 	\cr
\hline
\multicolumn{4}{c}{\textbf{FEROS}}	  \cr
\hline
Lines	&	Flux	&	Center	&	$\sigma$ 	\cr
& (arbitrary units) & (\AA) & (\AA) \cr
\hline
He~II$\lambda$6560	&	19.31 	&	6559.63 	&	0.49 	\cr
Raman He~II$\lambda$6545	&	2.14 	&	6545.80 	&	3.10 	\cr
He~II$\lambda$4859	&	2.70 	&	4859.00 	&	0.40 	\cr
Raman He~II$\lambda$4851	&	0.20 	&	4851.10 	&	2.00 	\cr
He~II$\lambda$4338	&	1.23 	&	4338.35 	&	0.39 	\cr
Raman He~II$\lambda$4332	&	0.09 	&	4332.36 	&	1.79 	\cr
\hline
 \end{tabular}
 \label{tab:FIT}
\end{table}

\subsection{Relative Velocities}

From the line-fitting analysis, we measured the line-center positions and fluxes of the Raman-scattered features and the nearby He~II emission lines. The line center positions provide kinematic information about the H~I region relative to the He~II emission region \citep[][]{choi20a,chang23}. The relative wavelength shift $\Delta \lambda_{c,A}$ and relative velocity $\Delta V_{c,A}$ between the H~I medium and He~II emission region are defined as follows:
\begin{equation}
   {\Delta \lambda_{c,A}} = (\lambda^{\rm atom}_{A} - \lambda^{\rm atom}_{RA}) - (\lambda^{\rm obs}_{A} - \lambda^{\rm obs}_{RA}),
    \label{equ:relative wavelength}
\end{equation}

\begin{equation}
   {\Delta V_{c,A} \over c} = \left(\frac{\lambda^{\rm atom}_{A'}}{\lambda^{\rm atom}_{RA}}\right) \ \frac{(\lambda^{\rm atom}_{A} - \lambda^{\rm atom}_{RA}) - (\lambda^{\rm obs}_{A} - \lambda^{\rm obs}_{RA}) }{\lambda^{\rm atom}_{RA}} ,
    \label{equ:relative}
\end{equation}
where $\lambda^{\rm obs}$ and $\lambda^{\rm atom}$ denote the observed and atomic line center wavelengths, respectively. Here, the subscript $A$ refers to optical He~II emission at 6560.10, 4859.32, and 4338.59 \AA, whereas the subscript $A'$ denotes ultraviolet He~II emission at 1025.28, 972.13, and 949.32 \AA. The subscript $RA$ stands for Raman He~II at 6544.70, 4851.30, and 4331.74 \AA.

\begin{table}

\caption{Relative velocities and Raman conversion efficiencies of three Raman-scattered \HeII\ lines of GHOST and FEROS spectra.
The second and third columns are the relative wavelength shift $\Delta \lambda_{c,A}$ and relative velocity $\Delta V_{c,A}$ between the H~I and He~II regions, which are defined in Equation (\ref{equ:relative wavelength}) and Equation (\ref{equ:relative}), respectively. The fourth column is the Raman conversion efficiency defined in Equation (\ref{equ:general}).
}
 \begin{tabular}{ccccc}
\hline
\multicolumn{4}{c}{\textbf{ GHOST}} \cr	
\hline
Raman-scattered \HeII\		&	$\Delta \lambda_{c,A}$	&	$\Delta V_{c,A}$ & RCE	\cr
  & (\AA) & (km \ s$^{-1}$) & (\%)  \cr
$\lambda$6545	&	 		2.21 	&	15.9  & 10.83	\cr
$\lambda$4851	&	 		0.31 	&	3.9  & 4.74	\cr
$\lambda$4332	&	 		1.24 	&	18.8 &  4.58	\cr\hline

\multicolumn{4}{c}{\textbf{FEROS}}	\cr		
\hline
Raman-scattered \HeII\	&	$\Delta \lambda_{c,A}$	&	$\Delta V_{c,A}$ & RCE	\cr
  & (\AA) & (km \ s$^{-1}$) & (\%) \cr
$\lambda$6545		&	1.57 	&	11.3 	& 13.74 \cr
$\lambda$4851	 	&	0.12 	&	1.5 	& 7.19 \cr
$\lambda$4332		&	0.86 	&	13.1 	& 6.23 \cr
									
\hline
 \end{tabular}
 \label{tab:VEL}
\end{table}

\subsection{Raman Conversion Efficiency}\label{sec:RCE}

As noted earlier, higher-level Raman-scattered \HeII\ lines have smaller cross sections. Therefore, the Raman conversion efficiency (RCE) for different Raman-scattered \HeII\ lines serves as a diagnostic of the neutral hydrogen region. Since RCE increases with increasing H~I column density (\NHI) and covering factor of the H~I region \citep{choi20a,lee19}, it serves as a tracer of the distribution of the scattering region for Raman-scattering (i.e., H~I region).
If RCEs of multiple Raman-scattered \HeII\ lines are observable, their ratio allows us to estimate \NHI \citep{chang23,lim25}. 

The RCE for Raman-scattered \HeII\ emission is defined as the photon number ratio between the incident and Raman-scattered radiation.
That is,
\begin{equation}\label{eq:RCE_basic}
    RCE_{\lambda_{\rm Ram}} = {\Phi_{\lambda_{\rm Ram}} \over
    \Phi_{\lambda_{\rm Inc}}},
\end{equation}
where $\Phi_{\lambda_{\rm Ram}}$ and $\Phi_{\lambda_{\rm Inc}}$ are the number fluxes of Raman-scattered He~II and the corresponding far-UV incident
He~II. For example, the RCE for Raman-scattered He~II$\lambda$6545 is defined as
\begin{equation}\label{eq:RCE1} 
{RCE}_{6545} = {{\Phi_{6545}} \over {\Phi_{1025}}} 
\end{equation}
where $\Phi_{1025}$ and $\Phi_{6545}$ denote the total photon fluxes of He~II$\lambda$1025 and Raman-scattered He~II$\lambda$6545, respectively.
Similarly, we define $RCE_{4851}$ and $RCE_{4332}$ for Raman-scattered He~II$\lambda$4851 and He~II$\lambda$4332, respectively. The RCE is expected to decrease for higher transitions at the same \NHI due to smaller Raman cross sections (see Table~\ref{tab:atomic}).

In estimating the RCEs of Raman-scattered He~II features, we must infer the number fluxes of the incident He~II, as they cannot be directly measured owing to strong interstellar extinction.
In particular, the He~II$\lambda$1025 line lies close to the \HI\ Lyman series, where extinction is extremely strong, rendering direct measurements impossible. A practical solution is to employ photoionization modeling, which yields the flux ratio between He~II$\lambda$1025 and an observable optical line such as He~II$\lambda$6560. By combining the observed flux of He~II$\lambda$6560 with the model-predicted ratio, $\Phi_{1025}$ can be derived indirectly. Consequently, Equation~(6) may be reformulated as
\begin{equation}
   RCE_{6545} = \alpha_{6545}\left( {F_{6545}\over F_{6560}} \right),
\end{equation}
where the coefficient $\alpha_{6545}$ is defined as
\begin{equation}
\alpha_{6545} = 
   \left(\frac{6545}{1025}\right)
   \left(\frac{F_{6560}}{F_{1025}}\right) \ .
\label{equ:alpha}
\end{equation}
In general, we may write
\begin{equation}
   RCE_{\lambda_{\rm Ram}} = \alpha_{\lambda_{\rm Ram}}\left( {F_{\lambda_{\rm Ram}}\over F_{\lambda_{\rm Optical}}} \right),
    \label{equ:general}
\end{equation}
yielding the RCE expressed in observable fluxes once the coefficients $\alpha_{\lambda_{\rm Ram}}$ are determined.

We employed the publicly available photoionization modeling code \texttt{CLOUDY} \citep{ferland17} to estimate the three coefficients $\alpha_{\lambda_{\rm Ram}}$.
In the same way as in \cite{lim25}, we assumed a spherical shell geometry to derive the relevant He~II fluxes. The nebula was modeled as having a uniform density, photoionized by a blackbody with $T_{\rm eff} = 2.2\times 10^5\,{\rm K}$ and $L = 1.43 \times 10^4 L_\odot$ \citep{santander_garcia17}, from which we found that
\begin{equation}
    \alpha_{6545}=1.24, \ 
    \alpha_{4851}=0.97, \ 
    \alpha_{4332}=0.87. \ 
\end{equation}

Table~\ref{tab:VEL} summarizes the relative wavelength shifts $\Delta \lambda_{c,A}$, the relative velocities $\Delta V_{c,A}$ between the H~I and He~II regions, and the values of the calculated RCEs for the three Raman-scattered He~II features obtained with GHOST and FEROS.

\section{Results}\label{sec:result}

We analyzed two high-resolution optical spectra of RR~Tel obtained roughly 20 years apart. 

We aim to obtain information on the changes in the distribution and kinematics of neutral hydrogen from the intensities and line centers of the Raman scattering lines in the two observational datasets.
In Section~\ref{sec:result_obs}, we describe and explore the differences in the Raman-scattered \HeII\ features between the two spectra based on our line-fitting analysis. In Section~\ref{sec:modeling}, we constrain the distribution of the H~I region by comparing the observed Raman-scattered \HeII\ lines with simulations from the 3D Monte Carlo radiative transfer code \texttt{STaRS} \citep{chang20}.

\subsection{Observed Properties of the Raman-scattered \HeII\ Lines}\label{sec:result_obs}

The relative velocities of the Raman-scattered \HeII\ lines, $\Delta V_{c,A}$, quantify the motion of the H~I scattering medium with respect to the \HeII\ emission region. In Table~\ref{tab:VEL}, all three Raman features show positive $\Delta V_{c,A}$ values in both the GHOST and FEROS spectra, indicating that the H~I medium, where \HeII\ UV photons are Raman-scattered, is outflowing relative to the \HeII-emitting region.

However, the relative velocities do not follow a monotonic trend across the three Raman transitions. In both spectra, Raman-scattered \HeII$\lambda$4851 exhibits a $\Delta V_{c,A}$ value that is $\approx 10~\rm km\,s^{-1}$ smaller than that of Raman-scattered \HeII$\lambda$6545, whereas Raman-scattered \HeII$\lambda$4332 shows the largest offset. Furthermore, the GHOST data display systematically larger $\Delta V_{c,A}$ values than the FEROS data. These trends imply that the kinematics of the neutral gas are not described by a single outflow speed but instead reflect a more complex velocity structure. We further discuss the non-monotonic behavior of the velocities in Section~\ref{sec:single_zone}.

Table~\ref{tab:VEL} also summarizes the RCEs for the three Raman-scattered \HeII\ lines. As expected from their respective Raman cross sections in Table~\ref{tab:atomic}, the efficiencies decrease in the sequence $RCE_{6545}$, $RCE_{4851}$, and $RCE_{4332}$ for both spectra. However, a significant temporal change is that the RCE values measured from the 2024 GHOST data are approximately 20--35\% lower than those from the 2004 FEROS data.
Because the RCE depends sensitively on the properties of the neutral region -- primarily the H~I column density $\NHI$ and the effective covering factor -- the observed differences likely reflect structural or dynamical changes in the H~I envelope over the past two decades, which also depend on the orbital phase. In the next section, we compare the measured RCE values with Monte Carlo radiative transfer simulations to constrain the distribution and evolution of the neutral hydrogen.

\subsection{Modeling the H~I Distribution with RCEs}\label{sec:modeling}

\begin{figure*}
    \centering
    \includegraphics[width=1.0\linewidth]{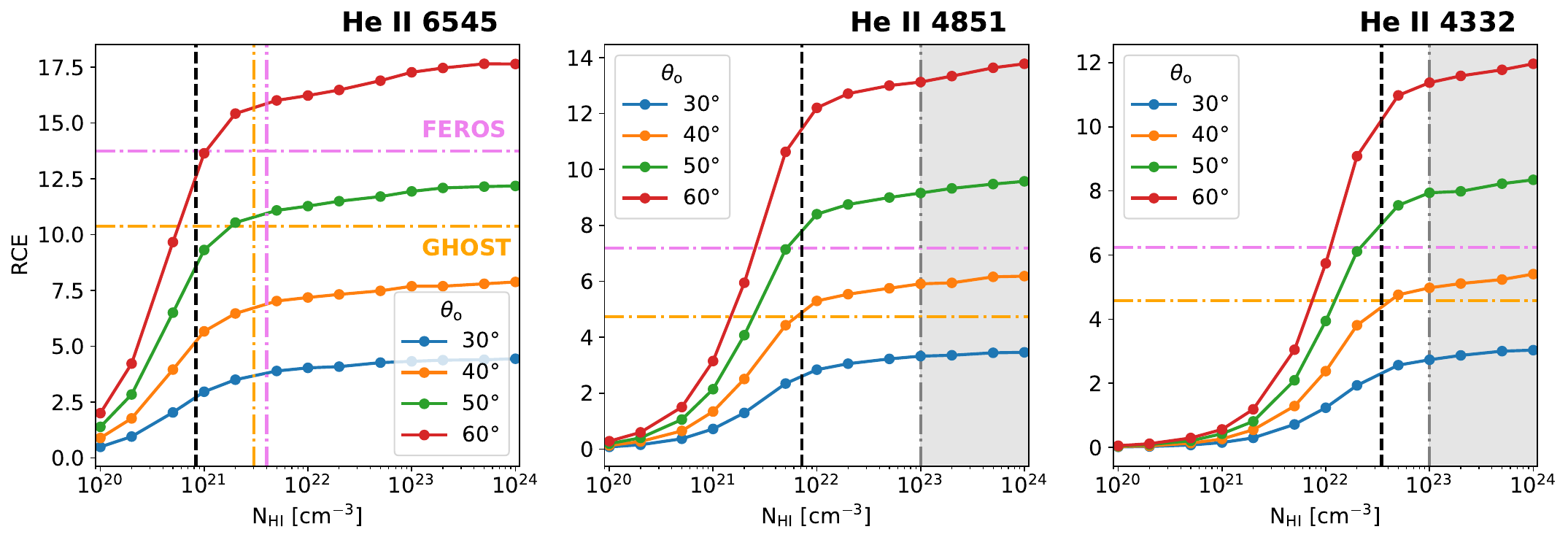}
    \caption{
    Simulated Raman conversion efficiencies (RCEs) for the Raman-scattered \HeII$\lambda$6545 (left), 4851 (center), and 4332 (right) features as a function of \NHI\ in the range of $10^{20}$--$10^{24}~\unitNHI$. 
    Solid curves indicate different opening angles $\theta_o$ of the H~I  region, ranging from $30^\circ$ to $60^\circ$.
    Horizontal dot–dashed lines denote the observed RCE values from the GHOST (orange) and FEROS (violet) spectra (Table~\ref{tab:VEL}). 
    Vertical black dashed lines represent the H~I column densities at which the Raman optical depth is unity, $\tau_{\rm Raman} = \NHI \sigma^{\rm Ram}_{2s} = 1$, for each line.
    In the left panel, 
    the vertical orange and violet dot–dashed lines indicate $\NHI \approx 3 \times 10^{21}\,\unitNHI$ (GHOST) and $4 \times 10^{21}\,\unitNHI$ (FEROS)
    estimated from the observed ratio $RCE_{6545}/RCE_{4851}$ in the left panel of Figure~\ref{fig:RCE_ratio}. 
    In the center and right panels, the gray dot–dashed line marks $\NHI \approx 1\times10^{23}~\unitNHI$ from the FEROS ratio $RCE_{4851}/RCE_{4332}$, and the gray shaded region indicates $\NHI > 10^{23}~\unitNHI$; the GHOST ratio falls below the values reproduced by the single-zone model and is therefore shown as a lower limit ($\NHI > 1\times10^{23}~\unitNHI$).
    The flattening of the RCE profiles to the right of the black dashed lines shows that the RCE becomes insensitive to increasing \NHI once the scattering region is optically thick ($\tau_{\rm Raman} > 1$).}
    \label{fig:RCE}
\end{figure*}

\begin{figure*}
    \centering
    \includegraphics[width=1.0\linewidth]{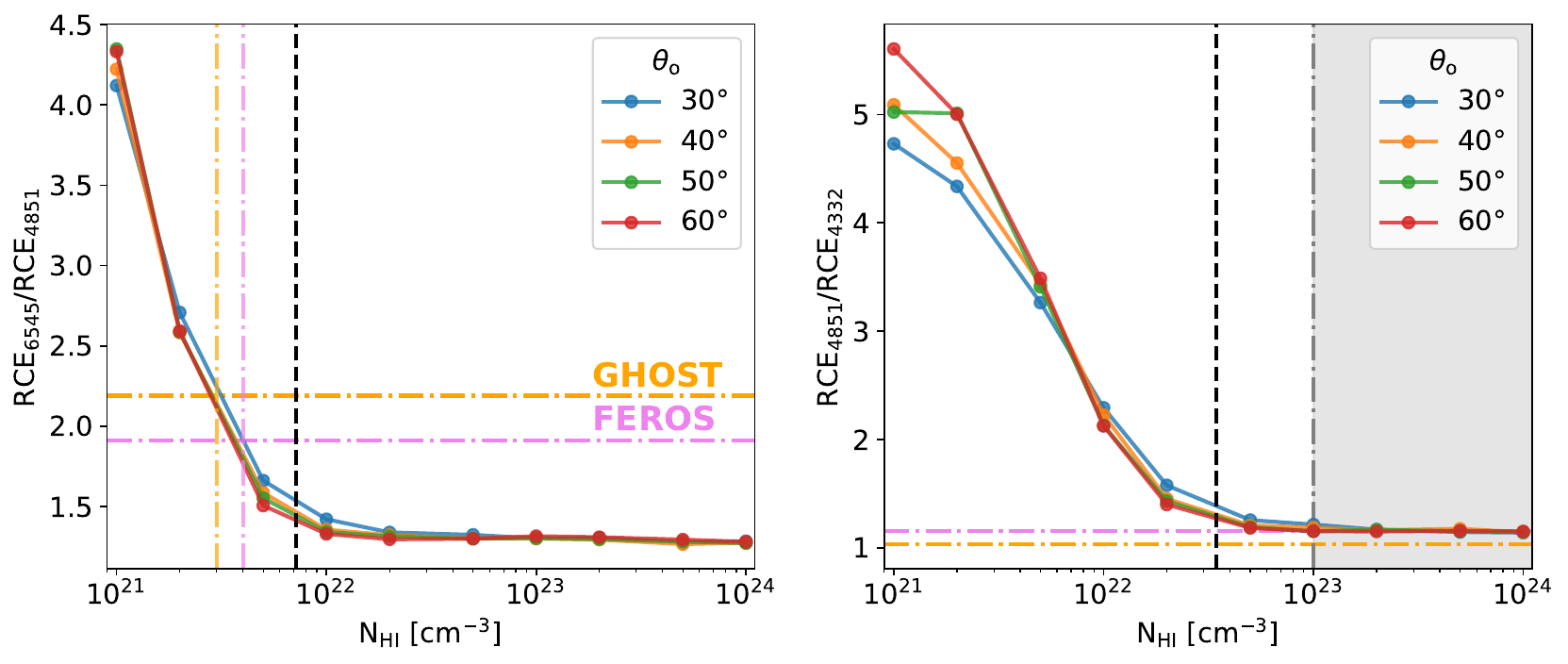}
    \caption{
    Ratios of the simulated Raman conversion efficiencies, $RCE_{6545}/RCE_{4851}$ (left) and $RCE_{4851}/RCE_{4332}$ (right), as functions of \NHI. 
    Horizontal dot–dashed lines represent the observed ratios from the GHOST (orange) and FEROS (violet) spectra as shown in Table~\ref{tab:VEL}. 
    Vertical black dashed lines in each panel mark the column density for which $\tau_{\rm Raman} = 1$ for the corresponding shorter-wavelength Raman-scattered line (\HeII$\lambda$4851 in the left panel and \HeII$\lambda$4332 in the right panel).
    In the left panel, the vertical orange and violet dot–dashed lines denote the \NHI values that reproduce the observed $RCE_{6545}/RCE_{4851}$ ratios for the GHOST ($\approx 3\times10^{21}~\unitNHI$) and FEROS ($\approx 4\times10^{21}~\unitNHI$) spectra.
    In the right panel, the gray dot–dashed line marks the $\NHI \approx 1\times10^{23}~\unitNHI$ that reproduces the observed FEROS ratio $RCE_{4851}/RCE_{4332}$, and the gray shaded region indicates $\NHI > 10^{23}~\unitNHI$; the GHOST ratio falls below the values reproduced by the single-zone model and is shown as a lower limit ($\NHI > 1\times10^{23}~\unitNHI$).
    Other information is the same as in Figure \ref{fig:RCE}.
    }
    \label{fig:RCE_ratio}
\end{figure*}

\begin{table}
\centering
\caption{
Estimated H~I column densities $\NHI$ and opening angles $\theta_o$ of the neutral scattering region, derived from the Raman conversion efficiencies of the Raman-scattered \HeII\ features at 6545, 4851, and 4332~\AA. 
The listed $\NHI$ and $\theta_o$ values correspond to those that reproduce the observed RCEs and RCE ratios shown in Figures~\ref{fig:RCE} and \ref{fig:RCE_ratio}.
}
\begin{tabular}{cccc}
\hline
RCE ratio & Data & $\NHI$ [\unitNHI] & $\theta_o$ \\
\hline
6545/4851   & GHOST& $3\times10^{21}$  & $50^\circ$ \\
            & FEROS & $4\times10^{21}$  & $55^\circ$ \\
4851/4332   &  GHOST & $>1\times10^{23}$  & $35^\circ$ \\
            & FEROS & $1\times10^{23}$  & $50^\circ$ \\
\hline
\end{tabular}
\label{tab:RCE}
\end{table}

To investigate the properties of the H~I region,
we use the 3D Monte Carlo radiative transfer code \texttt{STaRS} \citep[][]{chang20}.
We consider the geometry composed of a central \HeII\ point source surrounded by a neutral scattering medium.
The neutral region is a pure H~I medium characterized by two key parameters: an opening angle $\theta_o$ and an H~I column density \NHI. 
The central source isotropically emits the far-UV \HeII\ emission lines, which can be observed as the Raman-scattered \HeII\ lines at 6545, 4851, and 4332 \AA\ through Raman-scattering with atomic hydrogen.
Details of the adopted scattering geometry are provided in Appendix~\ref{sec:geometry}. To cover three Raman-scattered \HeII\ lines, we compute models with $\NHI$ in the range $10^{20}$--$10^{24}\,\unitNHI$.
Figures~\ref{fig:RCE} and \ref{fig:RCE_ratio} present the resulting RCEs and RCE ratios as functions of $\theta_o$ and $\NHI$, enabling a direct comparison with the observed values.

In Figure~\ref{fig:RCE}, the RCEs increase with \NHI when the Raman optical depth, $\tau_{\rm Raman} = \NHI\sigma_{2s}^{\rm Ram}$, remains below unity. 
Once the scattering region becomes optically thick ($\tau_{\rm Raman} > 1$), the RCE curves flatten because additional increases of \NHI no longer enhance the number of Raman-scattered photons; the incident UV photons are already almost fully converted through multiple scatterings within the optically thick medium. 
Thus, the three Raman-scattered \HeII\ lines are most sensitive to the regime in which the corresponding H~I layer is optically thin (i.e., $\NHI \lesssim 1/\sigma_{2s}^{\rm Ram}$). 
The RCEs also increase monotonically with the opening angle $\theta_o$, since a larger $\theta_o$ increases the fraction of incident photons intercepted by neutral gas. 
Because the covering factor scales as $\theta_o/180^\circ$, dividing one RCE by another removes this geometric dependence. 
Consequently, the ratios of RCEs for different Raman-scattered \HeII\ transitions are independent of $\theta_o$.

In Figure~\ref{fig:RCE_ratio}, we find that both $RCE_{6545}/RCE_{4851}$ and $RCE_{4851}/RCE_{4332}$ decrease with increasing \NHI, and their profiles do not depend on $\theta_o$. 
In the optically thin regime ($\tau_{\rm Raman} < 1$) for both lines, the ratios approach the ratios of their Raman-scattering cross sections. 
In contrast, once both transitions are optically thick ($\tau_{\rm Raman} > 1$), their RCEs saturate, and the ratios converge to constant values \citep{lee2016,chang23}. 
This saturation produces the flat portions of the curves in Figure~\ref{fig:RCE_ratio}, characteristic of the fully optically thick regime.

Because the RCE ratios are independent of $\theta_o$, they directly constrain \NHI of the scattering medium (e.g., \citealt{chang23,lim25}). 
In the left panel of Figure~\ref{fig:RCE_ratio}, the observed ratios $RCE_{6545}/RCE_{4851}$ are 
 2.28 (GHOST) and 1.91 (FEROS), corresponding to $\NHI \approx 3 \times 10^{21}~\unitNHI$ (GHOST) and $\approx 4 \times 10^{21}~\unitNHI$ (FEROS).
In contrast, the right panel shows that the observed ratios $RCE_{4851}/RCE_{4332}$ -- 1.03 for GHOST and 1.15 for FEROS -- correspond to substantially higher column densities of $\NHI \approx 1\times10^{23}~\unitNHI$ (FEROS) and a lower limit $\NHI > 1\times10^{23}~\unitNHI$ (GHOST, whose ratio falls below the values reproduced by the single-zone model).
Thus, for both data, \NHI inferred from $RCE_{6545}/RCE_{4851}$ are more than an order of magnitude smaller than those inferred from $RCE_{4851}/RCE_{4332}$.
This discrepancy demonstrates that a single value of \NHI cannot simultaneously reproduce all three Raman-scattered \HeII\ features, implying that the neutral region in RR~Tel must have a complex structure. 
Since our modeling assumes a single-zone H~I distribution, it is unable to capture this complexity; we further discuss this limitation in Section~\ref{sec:single_zone}.

In addition to estimating \NHI, we also constrain the opening angle $\theta_o$. 
In the left panel of Figure~\ref{fig:RCE}, at the column densities inferred from $RCE_{6545}/RCE_{4851}$ ($\NHI \approx 3 \times 10^{21}~\unitNHI$ for GHOST and $4 \times 10^{21}~\unitNHI$ for FEROS), $\theta_o \approx 50^\circ$ for GHOST and $\theta_o \approx 55^\circ$ for FEROS. 
From the center and right panels, adopting the higher \NHI from $RCE_{4851}/RCE_{4332}$ yields $\theta_o \approx 35^\circ$ (GHOST) and $\theta_o \approx 50^\circ$ (FEROS).
Table~\ref{tab:RCE} summarizes estimated $\NHI$ and $\theta_o$ values. 
Overall, the GHOST data suggest narrower opening angles than the FEROS data.
The implications of these differences -- including their connection to the relative velocity shifts of the Raman lines -- are discussed in Section~\ref{sec:interpretation}.

In summary, the values of $\NHI$ and $\theta_o$ inferred from the comparison between observed and simulated RCEs point to a neutral region whose structure cannot be described by a single-zone geometry. 
The markedly different \NHI required to match the three Raman-scattered \HeII\ features implies that each transition probes a distinct depth within the H~I envelope, reflecting spatial variations in both density and kinematics. 
These results highlight the need for a more complex kinematics and distribution of the H~I medium in RR~Tel, which we discuss further in the following section.

\section{Discussion}\label{sec:discussion}

We identify significant differences in the RCEs between the GHOST and FEROS datasets, indicating substantial changes in the Raman-scattered \HeII\ lines and the physical properties of the neutral hydrogen region. Raman-scattered \HeII\ lines in the 2004 FEROS data show higher RCEs and a smaller relative velocity than those in the 2024 GHOST data.
Our radiative transfer modeling further indicates a significant change in the physical properties of the neutral hydrogen region between the two datasets. 
However, the modeling results of each spectrum yield different \NHI\ values, estimated from the ratios $RCE_{6545}/RCE_{4851}$ and $RCE_{4851}/RCE_{4332}$.
This indicates the presence of a complex H~I medium structure, because the adopted model assumes a single-zone geometry characterized by a single \NHI. In the following, we discuss the physical origins of the differences between the two spectra in 2004 and 2024, and how Raman-scattered \HeII\ traces the complex kinematics and distribution of the H~I region in RR~Tel.

\subsection{Physical changes of the H~I region in RR~Tel}\label{sec:interpretation}

Our principal result is the significant change in the physical properties of the neutral hydrogen region between the FEROS (2004) and GHOST (2024) datasets. 
We show that the neutral region in 2004 had a larger covering factor (opening angle) than in 2024, as shown in Section~\ref{sec:modeling}. Several physical mechanisms may account for these changes, including intrinsic variations of the Mira wind and changes in the binary geometry.

One possible explanation lies in variations in Mira’s mass-loss rate, which is closely linked to the extent and density of the surrounding neutral region \citep[e.g.,][]{kotnikkaruza01}.
In particular, the smaller inferred covering factor in the GHOST data indicates that the neutral hydrogen region was less extended during the GHOST epoch than during the FEROS epoch.
When considered together with the larger relative velocities of Raman-scattered \HeII\ features in the GHOST data compared to those of the FEROS data (see Table~\ref{tab:VEL}), this suggests a change in the kinematic and geometric structure of the Mira wind. 

Although no significant photometric variations are evident in the AAVSO monitoring data, it should be noted that the optical magnitude may be dominated primarily by strong ionization lines (e.g., \Ha\ and \OIII) rather than by the stellar continuum of the giant star. Therefore, continued spectroscopic and photometric monitoring is required to clarify the detailed mass-loss processes and/or the binary orbital motion of RR~Tel \citep[e.g.,][]{jurkic12}.

Another possible explanation is orbital motion. Orbital motion may have caused a substantial change in the covering factor of the neutral region relative to the \HeII\ emission region. If the orbit is highly eccentric, the separation between the white dwarf and Mira components could vary significantly, thereby influencing the inferred neutral hydrogen column density and the covering factor, and hence the overall RCE.
An observational hint was provided by \cite{schmid02}, who conducted spectropolarimetric monitoring of the position angle of polarized Raman-scattered O~VI. 
Over seven years of observations of RR~Tel, they measured a mean rotation rate of 1.3$^\circ$ yr$^{-1}$, corresponding to an orbital period of about 270 years. 
While the 270-year orbital period lacks corroboration from alternative methods and the orbital elements remain poorly constrained, the 20-year gap between observation periods likely accounts for the significant variation in the resulting RCE.
In addition, the fraction of the neutral region illuminated by the UV emission depends on the observer's line of sight, analogous to variations in the projected area in ellipsoidal variables. 
However, it is beyond the scope of this paper to quantify the effect of the binary phase.

\subsection{Limitations of the single-zone model}\label{sec:single_zone}

The H~I column density is one of the crucial parameters to determine properties of Raman-scattered features such as Raman-scattered \HeII\ and O~VI lines \citep{schmid94,skopal06, choi20a,heo21}.
For this reason, we consider a geometry for radiative transfer modeling composed of a central source surrounded by a partial spherical region characterized by a single H~I column density \NHI\ and opening angle $\theta_o$ (i.e., a single-zone H~I geometry).
As the ratios of RCEs monotonically increase with increasing \NHI, we estimate \NHI using two RCE ratios, $RCE_{6545}/RCE_{4851}$ and $RCE_{4851}/RCE_{4332}$.

However, the values of \NHI and $\theta_o$ inferred from these two ratios are inconsistent (Table~\ref{tab:RCE}); the two ratios imply systematically different values of \NHI\ and $\theta_o$. 
Furthermore, Raman-scattered \HeII\ lines exhibit different relative velocities; $\Delta V_{c,A}$ at 4851 \AA\ is $\approx$10 km s$^{-1}$ smaller than that at 4332 \AA.
This indicates that the different Raman-scattered \HeII\ lines are formed in physically distinct regions of the H~I envelope with different kinematics. 

The H~I distribution is determined by stellar wind from a red giant, as discussed in Section~\ref{sec:interpretation}. Hence, \NHI at the emission region near a white dwarf is not isotropic in various lines of sight \citep{lee97}. Furthermore, spectropolarimetric observations of Raman-scattered O~VI features by \cite{schmid94} reveal a wavelength-dependent position angle, indicating a complex Raman-scattering behavior in anisotropic H~I distribution \citep[e.g.,][]{harries96}.
Thus, the neutral scattering region in RR~Tel cannot be fully described by a single-zone H~I geometry. Instead, the inferred $\NHI$ structure points to a stratified or multi-component neutral envelope with spatially varying density and kinematics.

\subsection{Raman He~II lines as tracers of complex H~I gas}

As shown in Section~\ref{sec:modeling}, RCE ratios can be used as indicators of the H~I column density. 
However, their interpretation is limited by the complexity of the H~I distribution, as discussed in Section~\ref{sec:single_zone}. 
One possible systematic uncertainty is the estimation of the incident far-UV \HeII\ photon fluxes, which are not directly observable and are instead inferred from the optical \HeII\ recombination lines using photoionization modeling. 
In this work, we adopted a fixed ionizing-source temperature and luminosity to compute the conversion factors $\alpha_{\lambda_{\rm Ram}}$. 
These assumptions may affect the ionization structure and the location of the H~I ionization front, and therefore can influence the quantitative values of \NHI\ and $\theta_o$ inferred from the RCEs.

However, the relative strengths of the \HeII recombination lines are not expected to vary strongly with the absolute luminosity of the ionizing source, and their dependence on the adopted source temperature is limited once the emitting gas remains in a highly ionized recombination regime \citep{osterbrock89, lim25}. 
Therefore, plausible variations in the ionizing-source parameters may shift the quantitative estimates of \NHI and $\theta_o$ of H I gas, but are unlikely to by themselves produce the observed line-dependent discrepancies. 
In particular, they do not explain the distinct relative velocity shifts among the Raman-scattered \HeII features. 
This suggests that the inconsistency between the RCE ratios is more naturally interpreted as evidence that the three Raman lines sample different regions of the neutral layers.

For this reason, we investigate the spatial distribution of locations where \HeII\ UV photons undergo Raman scattering (i.e., the last-scattering locations). Most Raman scattering occurs near the ionization front in optically thick H~I gas, whereas in the optically thin case, the scattering is more uniformly distributed (see Appendix~\ref{sec:spatial} for details). Consequently, even for media with the same \NHI, photons of each \HeII\ UV line are Raman-scattered at different locations due to their different cross sections, giving rise to a stratified spatial distribution of the Raman-scattered \HeII\ $\lambda$6545, 4851, and 4332 features.

As shown in Table~\ref{tab:VEL}, the observed kinematic hierarchy is consistent with the stratified scattering picture inferred from the Monte Carlo simulations. This kinematic difference suggests that the kinematics of the neutral medium are coupled to the distribution of H~I. The representative outflow speed of the Mira wind is $10$–$20{\rm\  km \ s^{-1}}$, implying that the relative velocity of the main wind region is expected to be smaller than that of the launching site by this characteristic range. Indeed, the systematically smaller velocity of Raman \HeII$\lambda$4851 compared to \HeII$\lambda$4332 is evident in both the FEROS and GHOST spectra. Moreover, the overall relative velocities measured in 2004 are smaller than those in 2024 by $\sim 2-6 {\rm\ km\ s^{-1}}$.

The observed differences among the three Raman-scattered \HeII\ lines therefore suggest that they do not simply provide redundant measurements of a single H~I column density. 
Instead, each line carries complementary information about a different effective scattering depth and velocity component within the neutral envelope. 
This makes the Raman \HeII\ triplet a potentially powerful diagnostic of the stratified H~I wind in RR~Tel and other symbiotic stars.



\subsection{Dusty neutral region}\label{sec:dusty_medium}

The IR excess near $3{\rm \ \mu m}$ shown in RR~Tel indicates the presence of a dust component with a temperature $T\sim 10^3{\rm\ K}$ \citep[e.g.,][]{angeloni10, jurkic12}.
In understanding the discrepancy between the observations and the theoretical model, an additional factor that warrants consideration is the possible influence of dust within the neutral scattering region.

The scattering cross sections for the far-UV \HeII\ lines at 1025, 972, and 949 \AA\ differ substantially, and therefore the formation sites of the corresponding Raman-scattered features are not identical, but instead are stratified according to the local H~I column density. 
For instance, the Raman-scattered \HeII$\lambda$4332 feature is expected to arise primarily in the deeper layers of the neutral region, where the column density reaches $N_{\rm HI}\sim10^{23}{\rm\ cm^{-2}}$. 
In contrast, the Raman-scattered feature at 6545 \AA\ originates in a relatively shallow layer characterized by a column density more than an order of magnitude lower.

If dust is present and mixed with neutral gas in these regions, extinction will preferentially affect shorter-wavelength photons; in particular, \HeII$\lambda$949 photons are much more susceptible to extinction than \HeII$\lambda$1025 photons \citep[e.g.,][]{seon16, lin24}.
This differential extinction would naturally lead to a selective suppression of the Raman-scattered \HeII$\lambda$4332 feature relative to the \HeII$\lambda$6545 feature. 
Consequently, variations in the dust content or its spatial distribution across the neutral region could provide an additional mechanism for the observed differences in RCE among the Raman features, complementing the role of geometry, kinematics, and ionization structure.\\

\subsection{Future modeling and observations}

A comprehensive radiative-transfer model is required to fully interpret the information imprinted on Raman-scattered \HeII\ lines. 
The present results show that the RCEs and velocity shifts cannot be explained by a single parameter, such as \NHI, or by a simple kinematic prescription alone. 
Instead, they are affected by several coupled factors, including the geometry and kinematics of the neutral wind, the ionization structure set by the white dwarf radiation field, and possible dust extinction within the neutral region.

Future modeling should therefore combine Raman radiative transfer with a physically motivated H~I distribution. 
This may be achieved by adopting a stellar-wind geometry \citep{lee97} or by using realistic hydrodynamic simulations \citep{lee22}. 
Realistic photoionization modeling is also required to estimate the \HeII\ emissivity and the location of the hydrogen ionization front, since the luminosity and temperature of the white dwarf, as well as the accretion rate, determine the ionization state of the gas in symbiotic stars \citep{Kuuttila2021a,Kuuttilab}. 
In addition, as discussed in Section~\ref{sec:dusty_medium}, dust in the neutral gas should be included, because differential far-UV extinction can modify the relative strengths of the incident \HeII\ lines before Raman scattering occurs.

Such modeling will inevitably involve degeneracies among wind geometry, ionization structure, and dust attenuation. 
Future high-resolution monitoring of the Raman-scattered \HeII{} features, particularly their temporal velocity shifts and profile asymmetries, can help constrain the relative contributions of different regions of the neutral wind.
Such observations can also test whether changes in the RCEs are mainly driven by variations in the neutral wind geometry, the ionization structure, or dust attenuation. 
A simultaneous analysis of the three Raman-scattered \HeII\ lines may therefore provide important constraints on the structure and kinematics of the slow stellar wind from the Mira component.

\section{Conclusion}

We have analyzed Raman-scattered \HeII\ lines using high-resolution spectra obtained with FEROS in 2004 and GHOST in 2024. By comparing the Raman conversion efficiencies over $\sim$20 yr baseline with Monte Carlo radiative transfer modeling, we investigated the structure and evolution of the neutral hydrogen scattering region. Our main conclusions are summarized as follows.

\begin{itemize}
    \item We detected the Raman-scattered \HeII\ lines blueward of \Ha, \Hb, and \Hg\ in both datasets, and performed line-fitting analysis to derive flux ratios relative to the neighboring optical \HeII\ emission lines.

    \item From these measurements, we derived the Raman conversion efficiencies (RCEs) and found that the RCEs in the 2004 FEROS data are substantially higher than those in the 2024 GHOST data.
    
    \item When interpreted with Monte Carlo modeling adopting a simple scattering geometry, these results suggest that the neutral region in 2004 had a larger covering factor (opening angle) than in 2024.

    \item However, the inconsistency between the RCEs inferred from different Raman \HeII\ lines demonstrates that the neutral scattering region cannot be fully described by a single-zone geometry.

    \item Together with distinct kinematics among the various Raman-scattered \HeII\ lines, this indicates that the Raman-scattered \HeII\ lines probe a stratified and kinematically complex neutral medium, making them a powerful diagnostic tool for studying the structure and kinematics of the neutral region in symbiotic systems.
    
\end{itemize}

Future work will involve radiative transfer modeling with refined scattering geometries that incorporate the effects of dust and binary orbital phase.

\section*{Acknowledgements}

This work was supported by the National Research Foundation of Korea (NRF) grants
funded by the Korea government (No. NRF-2023R1A2C1006984). This work was supported by the K-GMT Science Program (PID: GS-2024A-Q-305) of the Korea Astronomy and Space Science Institute (KASI).
Based on observations obtained at the international Gemini Observatory, a program of NSF NOIRLab, which is managed by the Association of Universities for Research in Astronomy (AURA) under a cooperative agreement with the U.S. National Science Foundation on behalf of the Gemini Observatory partnership: the U.S. National Science Foundation (United States), National Research Council (Canada), Agencia Nacional de Investigaci\'{o}n y Desarrollo (Chile), Ministerio de Ciencia, Tecnolog\'{i}a e Innovaci\'{o}n (Argentina), Minist\'{e}rio da Ci\^{e}ncia, Tecnologia, Inova\c{c}\~{o}es e Comunica\c{c}\~{o}es (Brazil), and Korea Astronomy and Space Science Institute (Republic of Korea). We acknowledge with thanks the variable star observations from the $AAVSO$ International Database contributed by observers worldwide and used in this research. We used generative AI tools (e.g., ChatGPT, OpenAI, and Claude, Anthropic) for language editing and coding assistance. All AI-generated suggestions were critically reviewed, tested, and validated by the authors, who take full responsibility for the scientific content and results.

\section*{Data Availability}

The data presented in this work are available through the Gemini Observatory Archive (Program ID: GS-2024A-Q-305; PI: Hee-Won Lee) and the ESO Science Archive (Program ID: 073.D-0724; PI: R. Zamanov). The STaRS code used for radiative transfer simulations is publicly available at https://github.com/csj607/STaRS.



\bibliographystyle{mnras}
\bibliography{example, references} 

@ARTICLE{Kuuttilab,
       author = {{Kuuttila}, J. and {Gilfanov}, M. and {Woods}, T.~E. and {Seitenzahl}, I.~R. and {Ruiter}, A.~J.},
        title = "{LIN 358: a symbiotic binary accreting above the steady hydrogen fusion limit}",
      journal = {\mnras},
         year = 2021,
        month = jan,
       volume = {500},
       number = {3},
        pages = {3763-3775},
          doi = {10.1093/mnras/staa3485},
archivePrefix = {arXiv},
       eprint = {2011.02864},
 primaryClass = {astro-ph.SR},
       adsurl = {https://ui.adsabs.harvard.edu/abs/2021MNRAS.500.3763K}
}

@ARTICLE{Kuuttila2021a,
       author = {{Kuuttila}, J. and {Gilfanov}, M.},
        title = "{Optical emission-line spectra of symbiotic binaries}",
      journal = {\mnras},
         year = 2021,
        month = oct,
       volume = {507},
       number = {1},
        pages = {594-607},
          doi = {10.1093/mnras/stab2025},
archivePrefix = {arXiv},
       eprint = {2107.07548},
 primaryClass = {astro-ph.SR},
       adsurl = {https://ui.adsabs.harvard.edu/abs/2021MNRAS.507..594K}
}

@ARTICLE{lee2016,
       author = {{Lee}, Young-Min and {Lee}, Dae-Sub and {Chang}, Seok-Jun and {Heo}, Jeong-Eun and {Lee}, Hee-Won and {Hwang}, Narae and {Park}, Byeong-Gon and {Lee}, Ho-Gyu},
        title = "{A Monte Carlo Study of Flux Ratios of Raman Scattered O VI Features at 6825 and 7082 {\r{A}} in Symbiotic Stars}",
      journal = {\apj},
         year = 2016,
        month = dec,
       volume = {833},
       number = {1},
          eid = {75},
        pages = {75},
          doi = {10.3847/1538-4357/833/1/75},
archivePrefix = {arXiv},
       eprint = {1610.07139},
 primaryClass = {astro-ph.SR},
       adsurl = {https://ui.adsabs.harvard.edu/abs/2016ApJ...833...75L}
}

@ARTICLE{Munari2019,
       author = {{Munari}, Ulisse},
        title = "{The Symbiotic Stars}",
      journal = {arXiv e-prints},
         year = 2019,
        month = sep,
          eid = {arXiv:1909.01389},
        pages = {arXiv:1909.01389},
          doi = {10.48550/arXiv.1909.01389},
archivePrefix = {arXiv},
       eprint = {1909.01389},
 primaryClass = {astro-ph.SR},
       adsurl = {https://ui.adsabs.harvard.edu/abs/2019arXiv190901389M}
}

@ARTICLE{Merc2025,
       author = {{Merc}, Jaroslav},
        title = "{Symbiotic Stars in the Era of Modern Ground- and Space-Based Surveys}",
      journal = {Galaxies},
         year = 2025,
        month = apr,
       volume = {13},
       number = {3},
          eid = {49},
        pages = {49},
          doi = {10.3390/galaxies13030049},
archivePrefix = {arXiv},
       eprint = {2504.16825},
 primaryClass = {astro-ph.SR},
       adsurl = {https://ui.adsabs.harvard.edu/abs/2025Galax..13...49M}
}

@ARTICLE{Placco2024,
       author = {{Placco}, Vinicius M. and {Herrera}, David and {Merino}, Brian M. and {US National Gemini Office} and {Hirst}, Paul and {Labrie}, Kathleen and {Simpson}, Chris and {Turner}, James and {Vacca}, William D. and {Gemini Science User Support Department} and {Deibert}, Emily and {Diaz}, Ruben and {Heo}, Jeong-Eun and {Kalari}, Venu and {Reggiani}, Henrique and {Rodriguez}, Cinthya and {Ruiz-Carmona}, Roque and {Thomas-Osip}, Joanna and {GHOST Instrument Team}},
        title = "{GHOST Reduced Data Products for the Gemini Observatory Community and Beyond}",
      journal = {Research Notes of the American Astronomical Society},
         year = 2024,
        month = dec,
       volume = {8},
       number = {12},
          eid = {312},
        pages = {312},
          doi = {10.3847/2515-5172/ad9f5f},
archivePrefix = {arXiv},
       eprint = {2412.13239},
 primaryClass = {astro-ph.IM},
       adsurl = {https://ui.adsabs.harvard.edu/abs/2024RNAAS...8..312P}
}

@ARTICLE{chang23,
       author = {{Chang}, Seok-Jun and {Lee}, Hee-Won and {Kim}, Jiyu and {Choi}, Yeon-Ho},
        title = "{Distribution and Kinematics of H I through Raman He II Spectroscopy of NGC 6302}",
      journal = {\apj},
         year = 2023,
        month = jun,
       volume = {949},
       number = {2},
          eid = {106},
        pages = {106},
          doi = {10.3847/1538-4357/acc868},
archivePrefix = {arXiv},
       eprint = {2303.16060},
 primaryClass = {astro-ph.SR},
       adsurl = {https://ui.adsabs.harvard.edu/abs/2023ApJ...949..106C}
}

@ARTICLE{cho10,
       author = {{Cho}, Se-Hyung and {Kim}, Jaeheon},
        title = "{Simultaneous Observations of SiO and H$_{2}$O Masers Toward Symbiotic Stars}",
      journal = {\apj},
         year = 2010,
        month = aug,
       volume = {719},
       number = {1},
        pages = {126-130},
          doi = {10.1088/0004-637X/719/1/126},
       adsurl = {https://ui.adsabs.harvard.edu/abs/2010ApJ...719..126C}
}

@ARTICLE{devalborro17,
       author = {{de Val-Borro}, M. and {Karovska}, M. and {Sasselov}, D.~D. and {Stone}, J.~M.},
        title = "{Three-dimensional hydrodynamical models of wind and outburst-related accretion in symbiotic systems}",
      journal = {\mnras},
         year = 2017,
        month = jul,
       volume = {468},
       number = {3},
        pages = {3408-3417},
          doi = {10.1093/mnras/stx684},
archivePrefix = {arXiv},
       eprint = {1704.03460},
 primaryClass = {astro-ph.SR},
       adsurl = {https://ui.adsabs.harvard.edu/abs/2017MNRAS.468.3408D}
}

@ARTICLE{akras19,
       author = {{Akras}, Stavros and {Guzman-Ramirez}, Lizette and
         {Leal-Ferreira}, Marcelo L. and {Ramos-Larios}, Gerardo},
        title = "{A Census of Symbiotic Stars in the 2MASS, WISE, and Gaia Surveys}",
      journal = {\apjs},
         year = 2019,
        month = feb,
       volume = {240},
       number = {2},
          eid = {21},
        pages = {21},
          doi = {10.3847/1538-4365/aaf88c},
archivePrefix = {arXiv},
       eprint = {1902.01451},
 primaryClass = {astro-ph.SR},
       adsurl = {https://ui.adsabs.harvard.edu/abs/2019ApJS..240...21A}
}

@ARTICLE{angeloni10,
       author = {{Angeloni}, R. and {Contini}, M. and {Ciroi}, S. and {Rafanelli}, P.},
        title = "{The spectral energy distribution of D-type symbiotic stars: the role of dust shells}",
      journal = {\mnras},
         year = 2010,
        month = mar,
       volume = {402},
       number = {3},
        pages = {2075-2086},
          doi = {10.1111/j.1365-2966.2009.16067.x},
archivePrefix = {arXiv},
       eprint = {0911.4853},
 primaryClass = {astro-ph.SR},
       adsurl = {https://ui.adsabs.harvard.edu/abs/2010MNRAS.402.2075A}
}

@ARTICLE{belczynski00,
       author = {{Belczy{\'n}ski}, K. and {Miko{\l}ajewska}, J. and {Munari}, U. and {Ivison}, R.~J. and {Friedjung}, M.},
        title = "{A catalogue of symbiotic stars}",
      journal = {\aaps},
         year = 2000,
        month = nov,
       volume = {146},
        pages = {407-435},
          doi = {10.1051/aas:2000280},
archivePrefix = {arXiv},
       eprint = {astro-ph/0005547},
 primaryClass = {astro-ph},
       adsurl = {https://ui.adsabs.harvard.edu/abs/2000A&AS..146..407B}
}

@BOOK{bethe67,
       author = {{Bethe}, H. A. and {Salpeter}, E. E.},
        title = "{Quantum Mechanics of One and Two Electron Atoms}",
    publisher = {Berlin: Springer; New York: Academic Press},
	 year = "1957"
}

@ARTICLE{birriel04,
       author = {{Birriel}, J.~J.},
        title = "{Raman-Scattered He II at 4851 {\r{A}} in the Symbiotic Stars HM Sagittae and V1016 Cygni}",
      journal = {\apj},
         year = 2004,
        month = sep,
       volume = {612},
       number = {2},
        pages = {1136-1139},
          doi = {10.1086/422835},
       adsurl = {https://ui.adsabs.harvard.edu/abs/2004ApJ...612.1136B}
}

@ARTICLE{chang18,
       author = {{Chang}, Seok-Jun and {Lee}, Hee-Won and {Lee}, Ho-Gyu and
         {Hwang}, Narae and {Ahn}, Sang-Hyeon and {Park}, Byeong-Gon},
        title = "{Broad Wings around H{\ensuremath{\alpha}} and H{\ensuremath{\beta}} in the Two S-type Symbiotic Stars Z Andromedae and AG Draconis}",
      journal = {\apj},
         year = 2018,
        month = oct,
       volume = {866},
       number = {2},
          eid = {129},
        pages = {129},
          doi = {10.3847/1538-4357/aadf88},
archivePrefix = {arXiv},
       eprint = {1809.01799},
 primaryClass = {astro-ph.SR},
       adsurl = {https://ui.adsabs.harvard.edu/abs/2018ApJ...866..129C}
}

@ARTICLE{chang20,
       author = {{Chang}, Seok-Jun and {Lee}, Hee-Won},
        title = "{STaRS: A 3D Grid-based Monte Carlo Code for Radiative Transfer through Raman and Rayleigh Scattering with Atomic Hydrogen}",
      journal = {Journal of Korean Astronomical Society},
         year = 2020,
        month = dec,
       volume = {53},
        pages = {169-179},
          doi = {10.5303/JKAS.2020.53.6.169},
archivePrefix = {arXiv},
       eprint = {2012.03424},
 primaryClass = {astro-ph.SR},
       adsurl = {https://ui.adsabs.harvard.edu/abs/2020JKAS...53..169C}
}

@ARTICLE{choi20a, 
       author = {{Choi}, Bo-Eun and {Chang}, Seok-Jun and {Lee}, Ho-Gyu and
         {Lee}, Hee-Won},
        title = "{Line Formation of Raman-scattered He II {\ensuremath{\lambda}} 4851 in an Expanding Spherical H I Shell in Young Planetary Nebulae}",
      journal = {\apj},
         year = 2020,
        month = jan,
       volume = {889},
       number = {1},
          eid = {2},
        pages = {2},
          doi = {10.3847/1538-4357/ab61f9},
archivePrefix = {arXiv},
       eprint = {1912.06291},
 primaryClass = {astro-ph.SR},
       adsurl = {https://ui.adsabs.harvard.edu/abs/2020ApJ...889....2C}
}

@ARTICLE{harries96,
       author = {{Harries}, T.~J. and {Howarth}, I.~D.},
        title = "{Raman scattering in symbiotic stars. I. Spectropolarimetric observations.}",
      journal = {\aaps},
         year = 1996,
        month = oct,
       volume = {119},
        pages = {61-90},
       adsurl = {https://ui.adsabs.harvard.edu/abs/1996A&AS..119...61H}
}

@ARTICLE{heo21,
       author = {{Heo}, Jeong-Eun and {Lee}, Hee-Won and {Angeloni}, Rodolfo and {Palma}, Tali and {Di Mille}, Francesco},
        title = "{Raman-scattered O VI Features in the Symbiotic Nova RR Telescopii}",
      journal = {\apj},
         year = 2021,
        month = jul,
       volume = {915},
       number = {2},
          eid = {105},
        pages = {105},
          doi = {10.3847/1538-4357/ac03b1},
archivePrefix = {arXiv},
       eprint = {2105.09442},
 primaryClass = {astro-ph.SR},
       adsurl = {https://ui.adsabs.harvard.edu/abs/2021ApJ...915..105H}
}

@ARTICLE{jurkic12,
       author = {{Jurkic}, T. and {Kotnik-Karuza}, D.},
        title = "{Modelling of dust around the symbiotic Mira RR Telescopii during obscuration epochs}",
      journal = {\aap},
         year = 2012,
        month = aug,
       volume = {544},
          eid = {A35},
        pages = {A35},
          doi = {10.1051/0004-6361/201218776},
       adsurl = {https://ui.adsabs.harvard.edu/abs/2012A&A...544A..35J}
}

@BOOK{kenyon86,
       author = {{Kenyon}, S.~J.},
        title = "{The symbiotic stars}",
         year = 1986,
       adsurl = {https://ui.adsabs.harvard.edu/abs/1986syst.book.....K}
}

@ARTICLE{kokubo24,
       author = {{Kokubo}, Mitsuru},
        title = "{Rayleigh and Raman scattering cross-sections and phase matrices of the ground-state hydrogen atom, and their astrophysical implications}",
      journal = {\mnras},
         year = 2024,
        month = apr,
       volume = {529},
       number = {3},
        pages = {2131-2149},
          doi = {10.1093/mnras/stae515},
archivePrefix = {arXiv},
       eprint = {2308.04959},
 primaryClass = {astro-ph.HE},
       adsurl = {https://ui.adsabs.harvard.edu/abs/2024MNRAS.529.2131K}
}

@ARTICLE{kotnikkaruza01,
       author = {{Kotnik-Karuza}, D. and {Friedjung}, M.},
        title = "{RR Tel: Mass Loss Rate of the Cool Component}",
      journal = {Odessa Astronomical Publications},
         year = 2001,
        month = dec,
       volume = {14},
        pages = {44-46},
       adsurl = {https://ui.adsabs.harvard.edu/abs/2001OAP....14...44K}
}

@ARTICLE{lee97,
       author = {{Lee}, H. -W. and {Lee}, K.~W.},
        title = "{On the profiles and the polarization of Raman-scattered emission lines in symbiotic stars}",
      journal = {\mnras},
         year = 1997,
        month = may,
       volume = {287},
       number = {1},
        pages = {211-220},
          doi = {10.1093/mnras/287.1.211},
       adsurl = {https://ui.adsabs.harvard.edu/abs/1997MNRAS.287..211L}
}

@ARTICLE{lee03,
       author = {{Lee}, Hee-Won and {Sohn}, Young-Jong and {Kang}, Young Woon and {Kim}, Ho-Il},
        title = "{Raman-scattered He II {\ensuremath{\lambda}}6545 Line in the Symbiotic Star V1016 Cygni}",
      journal = {\apj},
         year = 2003,
        month = nov,
       volume = {598},
       number = {1},
        pages = {553-559},
          doi = {10.1086/378886},
archivePrefix = {arXiv},
       eprint = {astro-ph/0308084},
 primaryClass = {astro-ph},
       adsurl = {https://ui.adsabs.harvard.edu/abs/2003ApJ...598..553L}
}

@ARTICLE{lee01,
       author = {{Lee}, Hee-Won and {Kang}, Yong-Woo and {Byun}, Yong-Ik},
        title = "{Raman-scattered He II Line in the Planetary Nebula M2-9 and in the Symbiotic Stars RR Telescopii and HE 2-106}",
      journal = {\apjl},
         year = 2001,
        month = apr,
       volume = {551},
       number = {1},
        pages = {L121-L124},
          doi = {10.1086/319830},
archivePrefix = {arXiv},
       eprint = {astro-ph/0103032},
 primaryClass = {astro-ph},
       adsurl = {https://ui.adsabs.harvard.edu/abs/2001ApJ...551L.121L}
}

@ARTICLE{lee00,
       author = {{Lee}, Hee-Won and {Hyung}, Siek},
        title = "{Broad H{\ensuremath{\alpha}} Wing Formation in the Planetary Nebula IC 4997}",
      journal = {\apjl},
         year = 2000,
        month = feb,
       volume = {530},
       number = {1},
        pages = {L49-L52},
          doi = {10.1086/312479},
archivePrefix = {arXiv},
       eprint = {astro-ph/9911422},
 primaryClass = {astro-ph},
       adsurl = {https://ui.adsabs.harvard.edu/abs/2000ApJ...530L..49L}
}

@ARTICLE{lee12,
       author = {{Lee}, Hee-Won},
        title = "{Raman Scattered He II {\ensuremath{\lambda}}4332 in the Symbiotic Star V1016 Cygni}",
      journal = {\apj},
         year = 2012,
        month = may,
       volume = {750},
       number = {2},
          eid = {127},
        pages = {127},
          doi = {10.1088/0004-637X/750/2/127},
       adsurl = {https://ui.adsabs.harvard.edu/abs/2012ApJ...750..127L}
}

@ARTICLE{lee19,
       author = {{Lee}, Young-Min and {Lee}, Hee-Won and {Lee}, Ho-Gyu and
         {Angeloni}, Rodolfo},
        title = "{Stellar-wind accretion and Raman-scattered O VI features in the symbiotic star AG Draconis}",
      journal = {\mnras},
         year = 2019,
        month = aug,
       volume = {487},
       number = {2},
        pages = {2166-2176},
          doi = {10.1093/mnras/stz1374},
archivePrefix = {arXiv},
       eprint = {1905.05993},
 primaryClass = {astro-ph.SR},
       adsurl = {https://ui.adsabs.harvard.edu/abs/2019MNRAS.487.2166L}
}

@ARTICLE{lee22,
       author = {{Lee}, Young-Min and {Kim}, Hyosun and {Lee}, Hee-Won},
        title = "{Formation of the Asymmetric Accretion Disk from Stellar Wind Accretion in an S-type Symbiotic Star}",
      journal = {\apj},
         year = 2022,
        month = jun,
       volume = {931},
       number = {2},
          eid = {142},
        pages = {142},
          doi = {10.3847/1538-4357/ac67d6},
archivePrefix = {arXiv},
       eprint = {2205.04758},
 primaryClass = {astro-ph.SR},
       adsurl = {https://ui.adsabs.harvard.edu/abs/2022ApJ...931..142L}
}

@ARTICLE{lim25,
       author = {{Lim}, Jin and {Chang}, Seok-Jun and {Shin}, Jaejin and {Lee}, Hee-Won and {Kim}, Jiyu and {Kim}, Hak-Sub and {Choi}, Bo-Eun and {Lee}, Ho-Gyu},
        title = "{High-resolution BOES Spectroscopy of Raman-scattered He II{\ensuremath{\lambda}}6545 in Young Planetary Nebulae}",
      journal = {\apj},
         year = 2025,
        month = feb,
       volume = {979},
       number = {2},
          eid = {124},
        pages = {124},
          doi = {10.3847/1538-4357/ada273},
archivePrefix = {arXiv},
       eprint = {2501.03558},
 primaryClass = {astro-ph.SR},
       adsurl = {https://ui.adsabs.harvard.edu/abs/2025ApJ...979..124L}
}

@ARTICLE{lin24,
       author = {{Lin}, Zesen and {Yan}, Renbin},
        title = "{Nebular dust attenuation with the Balmer and Paschen lines based on the MaNGA survey}",
      journal = {\aap},
         year = 2024,
        month = nov,
       volume = {691},
          eid = {A201},
        pages = {A201},
          doi = {10.1051/0004-6361/202451339},
archivePrefix = {arXiv},
       eprint = {2410.05067},
 primaryClass = {astro-ph.GA},
       adsurl = {https://ui.adsabs.harvard.edu/abs/2024A&A...691A.201L}
}

@ARTICLE{mayall49,
       author = {{Mayall}, Margaret Walton},
        title = "{Recent Variations of RR Telescopii}",
      journal = {Harvard College Observatory Bulletin},
         year = 1949,
        month = feb,
       volume = {919},
        pages = {15-17},
       adsurl = {https://ui.adsabs.harvard.edu/abs/1949BHarO.919...15M}
}

@ARTICLE{Ivison95,
       author = {{Ivison}, R.~J. and {Seaquist}, E.~R. and {Schwarz}, H.~E. and {Hughes}, D.~H. and {Bode}, M.~F.},
        title = "{Millimetre continuum emission from symbiotic stars - I. The measurements}",
      journal = {\mnras},
         year = 1995,
        month = mar,
       volume = {273},
       number = {2},
        pages = {517-527},
          doi = {10.1093/mnras/273.2.517},
       adsurl = {https://ui.adsabs.harvard.edu/abs/1995MNRAS.273..517I}
}

@ARTICLE{merc19a,
       author = {{Merc}, Jaroslav and {G{\'a}lis}, Rudolf and {Wolf}, Marek},
        title = "{First Release of the New Online Database of Symbiotic Variables}",
      journal = {Research Notes of the American Astronomical Society},
         year = 2019,
        month = feb,
       volume = {3},
       number = {2},
          eid = {28},
        pages = {28},
          doi = {10.3847/2515-5172/ab0429},
       adsurl = {https://ui.adsabs.harvard.edu/abs/2019RNAAS...3...28M}
}

@ARTICLE{merc19b,
       author = {{Merc}, J. and {G{\'a}lis}, R. and {Wolf}, M.},
        title = "{New online database of symbiotic variables: Symbiotics in X-rays}",
      journal = {Astronomische Nachrichten},
         year = 2019,
        month = aug,
       volume = {340},
       number = {7},
        pages = {598-606},
          doi = {10.1002/asna.201913662},
       adsurl = {https://ui.adsabs.harvard.edu/abs/2019AN....340..598M}
}

@ARTICLE{mikolajewska12,
       author = {{Miko{\l}ajewska}, J.},
        title = "{Symbiotic Stars: Observations Confront Theory}",
      journal = {Baltic Astronomy},
         year = 2012,
        month = jan,
       volume = {21},
        pages = {5-12},
          doi = {10.1515/astro-2017-0352},
archivePrefix = {arXiv},
       eprint = {1110.2361},
 primaryClass = {astro-ph.SR},
       adsurl = {https://ui.adsabs.harvard.edu/abs/2012BaltA..21....5M}
}

@ARTICLE{nussbaumer89,
       author = {{Nussbaumer}, H. and {Schmid}, H.~M. and {Vogel}, M.},
        title = "{Raman scattering as a diagnostic possibility in astrophysics.}",
      journal = {\aap},
         year = 1989,
        month = mar,
       volume = {211},
        pages = {L27-L30},
       adsurl = {https://ui.adsabs.harvard.edu/abs/1989A&A...211L..27N}
}

@BOOK{osterbrock89,
       author = {{Osterbrock}, Donald E.},
        title = "{Astrophysics of gaseous nebulae and active galactic nuclei}",
         year = 1989,
       adsurl = {https://ui.adsabs.harvard.edu/abs/1989agna.book.....O}
}

@BOOK{sakurai67,
       author = {{Sakurai}, J. J.},
        title = "{Advanced Quantum Mechanics}",
    publisher = {Massachusettes: Addison-Wesley Publishing Company},
	 year = "1967"
}

@ARTICLE{santander_garcia17,
       author = {{Santander-Garc{\'\i}a}, M. and {Bujarrabal}, V. and {Alcolea}, J. and {Castro-Carrizo}, A. and {S{\'a}nchez Contreras}, C. and {Quintana-Lacaci}, G. and {Corradi}, R.~L.~M. and {Neri}, R.},
        title = "{ALMA high spatial resolution observations of the dense molecular region of NGC 6302}",
      journal = {\aap},
         year = 2017,
        month = jan,
       volume = {597},
          eid = {A27},
        pages = {A27},
          doi = {10.1051/0004-6361/201629288},
archivePrefix = {arXiv},
       eprint = {1609.06455},
 primaryClass = {astro-ph.SR},
       adsurl = {https://ui.adsabs.harvard.edu/abs/2017A&A...597A..27S}
}

@ARTICLE{schmid89,
       author = {{Schmid}, H.~M.},
        title = "{Identification of the emission bands at lambda lambda 6830, 7088.}",
      journal = {\aap},
         year = 1989,
        month = mar,
       volume = {211},
        pages = {L31-L34},
       adsurl = {https://ui.adsabs.harvard.edu/abs/1989A&A...211L..31S}
}

@ARTICLE{schmid94,
       author = {{Schmid}, H.~M. and {Schild}, H.},
        title = "{Raman scattered emission lines in symbiotic stars : a spectropolarimetric survey.}",
      journal = {\aap},
         year = 1994,
        month = jan,
       volume = {281},
        pages = {145-160},
       adsurl = {https://ui.adsabs.harvard.edu/abs/1994A&A...281..145S}
}

@ARTICLE{schmid02,
       author = {{Schmid}, H.~M. and {Schild}, H.},
        title = "{Orbital motion in symbiotic Mira systems}",
      journal = {\aap},
         year = 2002,
        month = nov,
       volume = {395},
        pages = {117-127},
          doi = {10.1051/0004-6361:20021295},
       adsurl = {https://ui.adsabs.harvard.edu/abs/2002A&A...395..117S}
}

@ARTICLE{selvelli00,
       author = {{Selvelli}, P.~L. and {Bonifacio}, P.},
        title = "{First results of UVES at VLT: revisiting RR Tel}",
      journal = {\aap},
         year = 2000,
        month = dec,
       volume = {364},
        pages = {L1-L5},
          doi = {10.48550/arXiv.astro-ph/0008060},
archivePrefix = {arXiv},
       eprint = {astro-ph/0008060},
 primaryClass = {astro-ph},
       adsurl = {https://ui.adsabs.harvard.edu/abs/2000A&A...364L...1S}
}

@ARTICLE{seon16,
       author = {{Seon}, Kwang-Il and {Draine}, Bruce T.},
        title = "{Radiative Transfer Model of Dust Attenuation Curves in Clumpy, Galactic Environments}",
      journal = {\apj},
         year = 2016,
        month = dec,
       volume = {833},
       number = {2},
          eid = {201},
        pages = {201},
          doi = {10.3847/1538-4357/833/2/201},
archivePrefix = {arXiv},
       eprint = {1606.02030},
 primaryClass = {astro-ph.GA},
       adsurl = {https://ui.adsabs.harvard.edu/abs/2016ApJ...833..201S}
}

@ARTICLE{skopal06,
       author = {{Skopal}, A. and {Vittone}, A.~A. and {Errico}, L. and {Otsuka}, M. and {Tamura}, S. and {Wolf}, M. and {Elkin}, V.~G.},
        title = "{Structure of the hot object in the symbiotic prototype Z Andromedae during its 2000-03 active phase}",
      journal = {\aap},
         year = 2006,
        month = jul,
       volume = {453},
       number = {1},
        pages = {279-293},
          doi = {10.1051/0004-6361:20052917},
archivePrefix = {arXiv},
       eprint = {astro-ph/0603718},
 primaryClass = {astro-ph},
       adsurl = {https://ui.adsabs.harvard.edu/abs/2006A&A...453..279S}
}

@ARTICLE{skopal23,
       author = {{Skopal}, Augustin},
        title = "{The Emergence of a Neutral Wind Region in the Orbital Plane of Symbiotic Binaries during Their Outbursts}",
      journal = {\aj},
         year = 2023,
        month = jun,
       volume = {165},
       number = {6},
          eid = {258},
        pages = {258},
          doi = {10.3847/1538-3881/acd193},
archivePrefix = {arXiv},
       eprint = {2305.04220},
 primaryClass = {astro-ph.SR},
       adsurl = {https://ui.adsabs.harvard.edu/abs/2023AJ....165..258S}
}

@ARTICLE{vangroningen93,
       author = {{van Groningen}, E.},
        title = "{Further evidence for Raman scattering in RR Tel.}",
      journal = {\mnras},
         year = 1993,
        month = oct,
       volume = {264},
        pages = {975-979},
          doi = {10.1093/mnras/264.4.975},
       adsurl = {https://ui.adsabs.harvard.edu/abs/1993MNRAS.264..975V}
}

@ARTICLE{ferland17,
       author = {{Ferland}, G.~J. and {Chatzikos}, M. and {Guzm{\'a}n}, F. and {Lykins}, M.~L. and {van Hoof}, P.~A.~M. and {Williams}, R.~J.~R. and {Abel}, N.~P. and {Badnell}, N.~R. and {Keenan}, F.~P. and {Porter}, R.~L. and {Stancil}, P.~C.},
        title = "{The 2017 Release of Cloudy}",
      journal = {Revista Mexicana de Astronomía y Astrofísica},
         year = 2017,
        month = Oct,
       volume = {53},
       number = {2},
        pages = {60},
          doi = {10.48550/arXiv.1705.10877},
archivePrefix = {arXiv},
       eprint = {2017RMxAA..53..385F},
 primaryClass = {astro-ph.GA},
       adsurl = {https://ui.adsabs.harvard.edu/abs/2017RMxAA..53..385F}
}




\appendix

\renewcommand{\thefigure}{A\arabic{figure}}
\setcounter{figure}{0} 


\section{Scattering Geometry} \label{sec:geometry}

\begin{figure}
    \centering
    \includegraphics[width=0.91\linewidth]{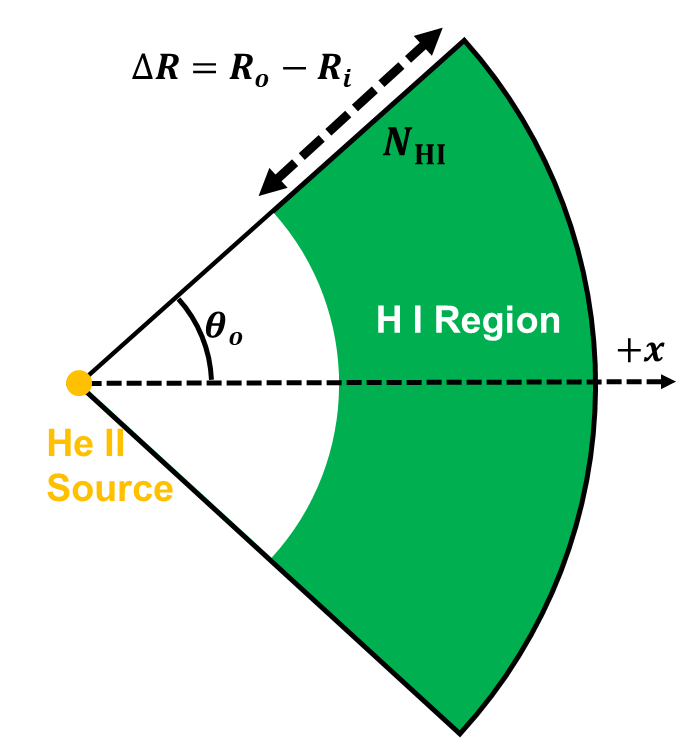}
    \caption{Schematic illustration of the scattering geometry for the radiative transfer modeling in Section~\ref{sec:modeling}. The geometry consists of a central \HeII\ emission source (orange) and a partial spherical H~I region (green).
    The central source emits UV \HeII\ lines at 1025, 972, and 949 \AA.
    The inner radius $R_i$ is half of the outer radius $R_o$.
    The H~I scattering region is characterized by a H~I column density \NHI.
    }
    \label{fig:geometry}
\end{figure}

In Section~\ref{sec:modeling}, we use the 3D grid-based radiative transfer code \texttt{STaRS} to compute the Raman conversion efficiency (RCE) of Raman \HeII\ lines at 6545, 4851, and 4332 \AA, assuming a single zone H~I column density. 
This section describes details of the scattering geometry and simulation.

Figure~\ref{fig:geometry} shows the schematic illustration of the scattering geometry, composed of a central \HeII\ emission region and H~I partial sphere.
The \HeII\ emission region is a highly ionized and compact gas near the white dwarf, while the H~I region from the red giant is spatially diffused and surrounds the ionized region.
As the emission region is much more compact than the H~I region, we set the \HeII\ emission source at the center of the geometry.
Since the neutral gas is coming from the direction of the red giant, we assume a partial spherical H~I geometry surrounding the central source in the side of the red giant ($+x$ direction).
The H~I region is characterized by an inner radius $R_i$, outer radius $R_o$, a half opening angle $\theta_o$, and H~I column density \NHI. As the inner region can be ionized by UV radiation from the white dwarf,  $R_i$ is fixed at $0.5 R_o$.
Therefore, we consider the broad ranges of $\theta_o = 30-60^\circ$ and $\NHI = 10^{20-23} \unitNHI$.

To compute the RCEs of three Raman-scattered \HeII\, 
The UV \HeII\ photons at 1025, 972, and 949 \AA\, are generated from the central source with the wavelength at the line center in our simulations; $10^5$ photons for each line.
After that, the simulations collect photons after they escape from the entire grid and estimate the ratio of Raman-scattered photons to the $n =2$ state, which is RCEs (see Eq.~\ref{eq:RCE_basic}). 
The detailed Monte Carlo method of \texttt{STaRS} is described in \cite{chang20}.

\section{Spatial distribution of Raman-scattering sites}\label{sec:spatial}
\renewcommand{\thefigure}{B\arabic{figure}}
\setcounter{figure}{0} 

We investigate the spatial distribution of Raman-scattering sites associated with the three Raman-scattered \HeII\ features. Because Raman optical photons are expected to escape the neutral region immediately after scattering, the location of these events provides a direct probe of the surface brightness distribution of the Raman-scattered emission. This approach, therefore, allows us to infer how the distribution of neutral hydrogen influences the emergent Raman features.  
If the scattering medium is optically thin, scattering takes place throughout the entire region with
almost equal probability, resulting in a uniform surface brightness. In contrast, for an optically thick scattering medium, scattering primarily occurs near the surface oriented toward the light source. 

\begin{figure}
    \centering
    \includegraphics[width=0.91\linewidth]{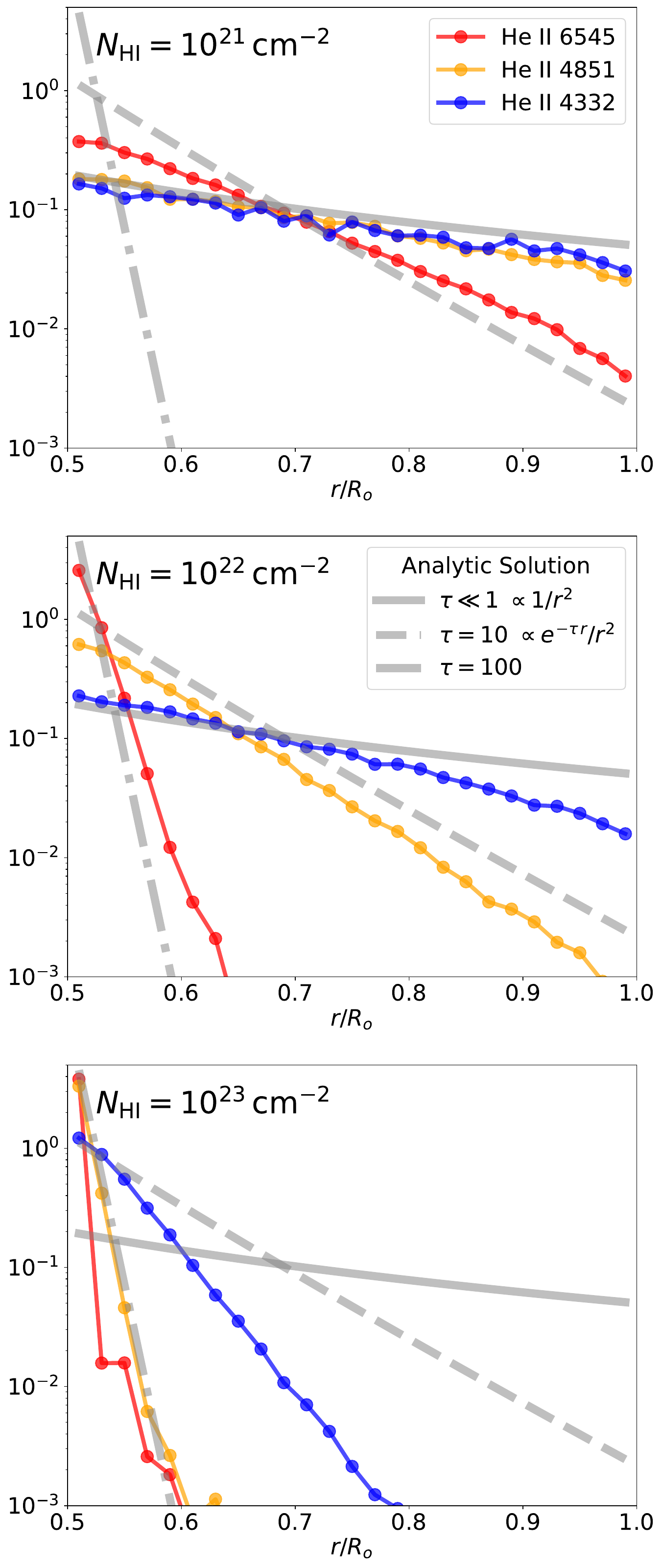}
    \caption{Histogram of the radius at the last-scattering location for different \NHI values of $10^{21}$ (top), $10^{22}$ (middle), and $10^{23} \unitNHI$ (bottom).
    The red, orange, and blue lines represent Raman-scattered \HeII$\lambda$6545, 4851, and 4332 features, respectively.
    The gray lines indicate the analytic solutions for three different optical depths: $\tau \ll 1$ (solid), $\tau = 10$ (dashed), and $\tau = 100$ (dot-dashed).
    }
    \label{fig:r_Raman}
\end{figure}

In Figure~\ref{fig:r_Raman}, we show the radial distribution of Raman photons measured from the center of the neutral region. The three panels correspond to column densities of $N_{\rm HI}=10^{21}$, $10^{22}$, and $10^{23}\ {\rm cm^{-2}}$, shown in the top, middle, and bottom panels, respectively. In the top panel, the scattering region is optically thin for all \HeII\ transitions, and the Raman-scattering that produces the three \HeII\ features is distributed nearly uniformly across the neutral region. In contrast, the bottom panel illustrates the optically thick case: the scattering region is opaque to Raman-scattered \HeII$\lambda$6545 and $\lambda$4851, and moderately thick for $\lambda$4332. Consequently, the Raman photons of \HeII$\lambda$6545 and $\lambda$4851 are confined within $\sim0.6,r/R_0$, whereas Raman \HeII$\lambda$4332 originates from deeper layers around $\sim0.8,r/R_0$.

In the middle panel, the medium is highly opaque to \HeII$\lambda$6545, moderately thick to \HeII$\lambda$4851, and optically thin to \HeII$\lambda$4332. At a column density of $N_{\rm HI}=10^{22}\ {\rm cm^{-2}}$, these three Raman-scattered \HeII\ features therefore exhibit markedly different surface brightness distributions.

\bsp	
\label{lastpage}
\end{document}